\documentclass[journal]{IEEEtran}

\usepackage{amsmath, amssymb, graphicx, hyperref, algorithm, algorithmic, subfigure, mathtools}

\usepackage{hyperref}
\hypersetup{
    colorlinks=true,     
    linkcolor=blue,       
    citecolor=blue,       
    urlcolor=black,        
}
\usepackage[table]{xcolor}

\newcounter{lemma} 

\newcounter{corollary} 

\newcounter{proposition} 

\newcounter{remarkjs} 

\def\ccalF{{\ensuremath{\mathcal F}}}

\def\ccalL{{\ensuremath{\mathcal L}}}

\def\ccalP{{\ensuremath{\mathcal P}}}

\def\ccal0{{\ensuremath{\mathcal 0}}}
\def\bbB{{\ensuremath{\mathbf B}}}

\def\bbD{{\ensuremath{\mathbf D}}}

\def\bbH{{\ensuremath{\mathbf H}}}
\def\bbI{{\ensuremath{\mathbf I}}}

\def\bbM{{\ensuremath{\mathbf M}}}

\def\bbO{{\ensuremath{\mathbf O}}}

\def\bbQ{{\ensuremath{\mathbf Q}}}

\def\bbW{{\ensuremath{\mathbf W}}}

\def\bbS{{\ensuremath{\mathbf S}}}
\def\bbT{{\ensuremath{\mathbf T}}}
\def\bbX{{\ensuremath{\mathbf X}}}
\def\bbY{{\ensuremath{\mathbf Y}}}
\def\bbZ{{\ensuremath{\mathbf Z}}}

\def\bbb{{\ensuremath{\mathbf b}}}

\def\bbd{{\ensuremath{\mathbf d}}}

\def\bbh{{\ensuremath{\mathbf h}}}

\def\bbp{{\ensuremath{\mathbf p}}}
\def\bbq{{\ensuremath{\mathbf q}}}

\def\bbw{{\ensuremath{\mathbf w}}}

\def\bbs{{\ensuremath{\mathbf s}}}

\def\bbx{{\ensuremath{\mathbf x}}}
\def\bby{{\ensuremath{\mathbf y}}}
\def\bbz{{\ensuremath{\mathbf z}}}
\def\bb0{{\ensuremath{\mathbf 0}}}
\def\bbx{{\ensuremath{\mathbf{x}}}}
\def\bbM{{\ensuremath{\mathbf{M}}}}
\def\bbD{{\ensuremath{\mathbf{D}}}}
\def\bbs{{\ensuremath{\mathbf{s}}}}
\def\bbW{{\ensuremath{\mathbf{W}}}}
\def\bbI{{\ensuremath{\mathbf{I}}}}
\def\bbX{{\ensuremath{\mathbf{X}}}}
\def\bbd{{\ensuremath{\mathbf{d}}}}

\def\c[#1]{{\ensuremath{^{(#1)}}}}

\newcommand{\dvec}{\mathbf{\textit{\textbf{d}}}}

\begin{document}

\title{Weakly Supervised Convolutional Dictionary Learning for Multi-Label Classification}


\author{
    Hao~Chen and
    Dayuan~Tan
    
    \thanks{H. Chen was with the Department of Electrical and Computer Engineering, University of Maryland, Baltimore County (\href{mailto:chenhao1@umbc.edu}{chenhao1@umbc.edu}).}%
    \thanks{D. Tan was with the Department of Electrical and Computer Engineering, University of Maryland, Baltimore County (\href{mailto:dayuan1@umbc.edu}{dayuan1@umbc.edu}).}
}


\maketitle


\begin{abstract}
    Convolutional Dictionary Learning (CDL) has emerged as a powerful approach for signal representation by learning translation-invariant features through convolution operations. While existing CDL methods are predominantly designed and used for fully supervised settings, many real-world classification tasks often rely on weakly labeled data, where only bag-level annotations are available.
    In this paper, we propose a novel weakly supervised convolutional dictionary learning framework that jointly learns shared and class-specific components, for multi-instance multi-label (MIML) classification where each example consists of multiple instances and may be associated with multiple labels. Our approach decomposes signals into background patterns captured by a shared dictionary and discriminative features encoded in class-specific dictionaries, with nuclear norm constraints preventing feature dilution.
    A Block Proximal Gradient method with Majorization (BPG-M) is developed  to alternately update dictionary atoms and sparse coefficients, ensuring convergence to local minima. Furthermore, we incorporate a projection mechanism that aggregates instance-level predictions to bag-level labels through learnable pooling operators.
    Experimental results on both synthetic and real-world datasets demonstrate that our framework outperforms existing MIML methods in terms of classification performance, particularly in low-label regimes. The learned dictionaries provide interpretable representations while effectively handling background noise and variable-length instances, making the method suitable for applications such as environmental sound classification and RF signal analysis.
\end{abstract}

\begin{IEEEkeywords}
        Dictionary learning, convolutional dictionary learning, weakly supervised learning, multi-instance multi-label learning, classification.
\end{IEEEkeywords}

\section{Introduction}
\label{sec:introduction}

Dictionary learning (DL) is a powerful technique for sparse representation, widely used in tasks such as denoising and data compression \cite{AEB06,BDE09,XY16}. There are two main types of dictionary learning: synthesis dictionary learning and analysis dictionary learning \cite{SPC14}. The former aims to represent data as the product of a dictionary and sparse coefficients, learning both components directly from the data. In contrast, the latter seeks to learn a projection operator that maps data into a sparse space. Both approaches are commonly used in supervised signal classification tasks, where label information is leveraged to guide signal reconstruction and modified Fisher criteria are used to regularize sparse coefficients for better discriminative ability \cite{FDDL, Tang19}.

Among recent developments, Convolutional Dictionary Learning (CDL) has emerged as a representative and effective variant of DL. It has shown its capability in image reconstruction with advantages such as translation invariance, and the ability to learn local features without redundancy. CDL methods have been extended in multiple directions, including distributed implementations (DiCoDiLe \cite{DiCoDiLe}), multi-modal extensions \cite{MultiModalCDL}, tensor formulations \cite{TensorCDL}, and integration with deep architectures (DeepM2CDL \cite{DeepM2CDL}). 
While several CDL approaches have been proposed for classification tasks \cite{Tang19, chen16, jin17, YRF18}, few have specifically addressed multi-instance multi-label (MIML) scenarios, even though such scenarios are prevalent in weakly supervised learning settings.

One reason for the limited exploration of MIML in CDL methods lies in the unique difficulties posed by the MIML framework itself. 
In MIML settings, each training example is represented as a bag of instances with a set of labels associated with the entire bag. However the precise labels of the individual instances within the bag remain unknown. This is common in domains like Radio Frequency (RF) or audio, where data is collected over a period of time and contain multiple signals, but the label of each signal is not revealed. For example, in a room with several Wi-Fi and Bluetooth transmitters, the collected data is a bag of instances with overall known labels (Wi-Fi and Bluetooth), yet it remains unclear which specific instances correspond to which labels. 

Existing CDL methods like DiCoDiLe \cite{DiCoDiLe} focus on distributed optimization but rely on the availability of instance-level labels, which make them unsuitable for weakly supervised MIML settings. Analysis CDL approaches such as \cite{YRF18} address weak supervision through probabilistic modeling but sacrifice interpretability and direct background modeling. The tensor formulation in \cite{TensorCDL} shows promise for handling complex data structures but fall short in addressing the fundamental challenges of label disambiguation in MIML settings.

Our objective is to design and develop a MIML classification algorithm that can predict the labels of a given bag. We propose a synthesis-based MIML-CDL framework that combines three key innovations:
\begin{itemize}
    \item It jointly learns shared background dictionary and class-specific dictionaries under nuclear norm constraints, thereby extending the multi-modal concepts in \cite{MultiModalCDL} to weakly supervised settings.
    \item It employs a projection mechanism that aggregates instance-level predictions via learnable pooling operators, inspired by deep aggregation in \cite{DeepM2CDL} but incorporates explicit dictionary constraints.
    \item It adapts Block Proximal Gradient method with Majorization (BPG-M) optimization for weakly supervised CDL, combining the convergence guarantees of \cite{ChunF18} with tensor-aware update strategies inspired by \cite{TensorCDL}.
\end{itemize}

Our experimental validation demonstrates that the proposed method achieves exceptional performance on both synthetic and real-world datasets compared to existing MIML methods, particularly in low-label regimes. The approach maintains interpretability through explicit dictionary learning while effectively addressing background noise and variable-length instances - challenges not adequately handled in distributed \cite{DiCoDiLe} or tensor \cite{TensorCDL} CDL formulations.

The paper is organized as follows. Section \ref{sec:related} reviews related work, and Section \ref{sec:background} provides necessary background. Section \ref{sec:problem} formalizes the MIML-CDL problem, while Section \ref{sec:method} details our proposed BPG-M optimization approach. Section \ref{sec:ext} discusses tensor extensions, and Section \ref{sec:results} presents comprehensive experimental evaluations. Finally, Section \ref{sec:conclusion} concludes the paper.
\section{Related Work}
\label{sec:related}

\subsection{Dictionary Learning and Convolutional Extensions}
Dictionary learning has evolved significantly since the seminal K-SVD algorithm \cite{AEB06}, which established the foundation for learning overcomplete representations from data. While early approaches focused on reconstruction quality, discriminative dictionary learning methods like FDDL \cite{FDDL, CHK18, CHK21} incorporated label information through Fisher discrimination criteria to enhance classification performance. Analysis dictionary learning \cite{SPC14} offered an alternative perspective by learning projections rather than decompositions.

CDL \cite{GRK07, HHW15} extended these concepts to handle translation invariance in signals, enabling more efficient representation of temporal and spatial patterns. Recent advances in CDL optimization have introduced distributed implementations (DiCoDiLe \cite{DiCoDiLe}), which enable scaling to large datasets but lack mechanisms for weak supervision. Our approach builds upon these foundations while introducing novel constraints specifically tailored for MIML scenarios.

\subsection{Weakly Supervised Learning in MIML Settings}
MIML learning \cite{ZZ2007} is designed for settings in which each training example consists of a bag of instances with only bag-level labels available. Traditional MIML methods, such as instance-space transformations \cite{MZ2010} and SVM adaptations \cite{CL2011}, require explicit identification of individual instances. More recent deep learning approaches like Deep MIML \cite{FZ17} have shown promise but typically require large labeled datasets and sacrifice interpretability.

Weakly supervised dictionary learning (WSDL) \cite{YRF18} represents an initial attempt to bridge the gap between dictionary learning and weak supervision. However, it primarily focuses on the analysis dictionary formulation and lacks mechanisms to handle background components or convolutional structures.

The separation of shared and discriminative features has proven effective in both traditional and deep learning contexts. For example, low-rank shared dictionary learning \cite{VuM17} demonstrated that explicitly modeling common patterns can improve classification performance by reducing feature redundancy. Multi-modal CDL approaches \cite{MultiModalCDL} have extended this concept to handle different data types; however, they typically assume strong supervision.

Our synthesis-based MIML-CDL framework addresses these limitations through explicit modeling of both shared and discriminative components. To address the challenges of handling the background noise, our approach introduces nuclear norm constraints on the shared dictionary $\bbD^{(0)}$ to prevent feature dilution, which is a critical consideration in weakly supervised scenarios where discriminative features might otherwise be absorbed into the background model. This extends the concepts from multi-modal learning \cite{MultiModalCDL} to the weakly supervised domain, addressing a significant gap in prior work.

\subsection{Optimization, Aggregation, and Applications}
Weakly supervised CDL problems present two major challenges: (1) efficiently optimizing non-convex objectives under dictionary constraints, and (2) aggregating instance-level representations into coherent bag-level decisions. In this section, we address each challenge in turn and propose novel solutions tailored to the multi-instance setting.

The computational challenges of CDL have inspired various optimization strategies. Among early efforts, Alternating Direction Method of Multipliers (ADMM)-based approaches \cite{BCE10, P19} played a key role in advancing CDL optimization, but they often struggle with convergence in non-convex settings. More recently, the Block Proximal Gradient method with Majorization (BPG-M) \cite{ChunF18} offers stronger theoretical guarantees for CDL problems. Tensor formulations \cite{TensorCDL} have further enhanced CDL by enabling structured low-rank constraints on activation tensors. 

Our adaptation of BPG-M for weakly supervised CDL combines the convergence guarantees of \cite{ChunF18} with tensor-aware updates inspired by \cite{TensorCDL}. This hybrid formulation enables efficient optimization of the non-convex objective while preserving model interpretability and scalability. 

Beyond optimization, a critical challenge in MIML learning is aggregating instance-level predictions to form bag-level decisions. Deep learning approaches like \cite{FZ17} typically use end-to-end differentiable pooling operations, while traditional methods often rely on simple heuristics like max or average pooling. Our approach introduces a learnable projection mechanism that aggregates instance-level sparse representations through adaptive pooling operators. Unlike deep aggregation techniques such as \cite{DeepM2CDL}, our method maintains explicit dictionary constraints throughout the aggregation process, which enhances interpretability while supporting flexible instance-to-bag mappings. This design strikes a thoughtful balance between purely data-driven deep approaches and rigid traditional pooling strategies.

Environmental sound classification presents particular challenges for weakly supervised methods due to temporal variability and background noise. While deep learning approaches \cite{KGN19} have achieved success with large-scale labeled datasets, our experiments demonstrate that structured dictionary learning can achieve competitive performance with greater interpretability, particularly in low-label regimes.

Our MIML-CDL framework integrates these research threads through three key innovations: (1) enforceing joint learning of shared and class-specific convolutional dictionaries with nuclear norm constraints, (2) introducing learnable projection mechanisms for instance-to-bag prediction aggregation, and (3) adapting BPG-M optimization techniques to accommodate weakly supervised settings. This combination addresses critical limitations in existing approaches while maintaining computational efficiency and model interpretability.
\section{Background} \label{sec:background}

\subsection{Dictionary Learning}
DL is a representation learning technique that aims to find a set of basis vectors (atoms) such that input signals can be represented as sparse linear combinations of these atoms. Formally, given a set of training signals $\mathbf{X} = [\mathbf{x}_1, \mathbf{x}_2, \ldots, \mathbf{x}_N] \in \mathbb{R}^{T \times N}$, dictionary learning seeks to find a dictionary $\mathbf{D} \in \mathbb{R}^{T \times K}$ and a sparse coefficient matrix $\mathbf{Z} \in \mathbb{R}^{K \times N}$ such that $\mathbf{X} \approx \mathbf{D}\mathbf{Z}$ and $\mathbf{Z}$ is sparse.

The standard synthesis dictionary learning problem can be formulated as:
\begin{equation} \label{eq:dl}
\min_{\mathbf{D}, \mathbf{Z}} \|\mathbf{X} - \mathbf{D}\mathbf{Z}\|_F^2 + \lambda\|\mathbf{Z}\|_1 \quad \text{subject to} \quad \|\mathbf{d}_k\|_2 \leq 1, \forall k
\end{equation}
where $\|\cdot\|_F$ denotes the Frobenius norm, $\|\cdot\|_2$ represents the $\ell_2$-norm, $\|\cdot\|_1$ is the $\ell_1$-norm that promotes sparsity, $\mathbf{d}_k$ is the $k$-th atom in the dictionary, and $\lambda$ is a regularization parameter that controls the trade-off between reconstruction error and sparsity.

\subsection{Convolutional Dictionary Learning}
CDL extends traditional dictionary learning by replacing matrix multiplication with convolution operations. This modification enables CDL to effectively model signals with spatial or temporal structure by capturing  local patterns that may appear at different locations within the signal.

In CDL, the reconstruction of any given signal $\mathbf{x}$ can be written as:
\begin{equation}
\mathbf{x} \approx \sum_{k=1}^{K} \mathbf{d}_k * \mathbf{z}_k
\end{equation}
where $*$ denotes the convolution operation, $\mathbf{d}_k$ represents the $k$-th atom in the dictionary, and $\mathbf{z}_k$ is its corresponding sparse feature map.

The key advantage of CDL over traditional DL is its translation invariance property, which allows it to detect similar patterns regardless of their position in the signal. This makes CDL particularly well-suited for tasks in image processing, audio analysis, and other domains where patterns may appear at various locations.

\subsection{Multi-Instance Multi-Label Learning}
MIML learning is a weakly supervised learning paradigm where each training example is represented as a bag of instances, and the bag is annotated with multiple labels. However, the instance-to-label relationships are not specified during training, which makes learning in the MIML setting both ambiguous and challenging.

Formally, in MIML, we have a training set $\{(\mathbf{X}_i, \mathbf{y}_i)\}_{i=1}^N$, where $\mathbf{X}_i = \{\mathbf{x}_{i_1}, \mathbf{x}_{i_2}, \ldots, \mathbf{x}_{i_m}\}$ is $i-th$ training example as a bag of $m$ instances, and $\mathbf{y}_i \in \{0,1\}^C$ is a binary vector indicating the presence or absence of $C$ possible labels associated with $\mathbf{X}_i$.

The MIML framework faces two main ambiguities:
\begin{itemize}
    \item Instance-label ambiguity: It is unknown which instances in a bag correspond to which labels.
    \item Label-correlation ambiguity: The model must account for possible interdependencies among labels, as they may not be mutually independent.
\end{itemize}

These ambiguities make MIML a complex yet practical solution for many real-world applications where obtaining fine-grained annotations is expensive or impractical. Examples include image classification where an image (bag) contains multiple objects (instances) and is associated with multiple tags (labels), or audio classification where a recording (bag) contains multiple sound events (instances) and is associated with multiple categories (labels).

\section{Problem Formulation}
\label{sec:problem}

\subsection{Dictionary Learning \& Convolutional Dictionary Learning}
\label{sec:DL_CDL}

In synthesis DL, as shown in Eq. \ref{eq:dl}, each input $\bbX$ is approximated as a linear combination of learned basis elements, which is represented by the product of a dictionary and sparse coefficients. Since standard synthesis DL focuses exclusively on reconstructing the data without using any label information, its basic form can be written as:
\begin{align}
    \min_{\bbD, \bbz_n} \sum_n^N \|\bbx_n - \bbD \bbz_n \|_2^2 + \lambda \|\bbz_n\|_1 ,\ \ \ \  \|\bbd_k\|\leq 1,
\end{align}
where $\bbD$ is a dictionary consisting of atoms that represent all possible features of the data, and $\lambda \| \cdot\|_1$ is the sparse regularizer. $\|\bbd_k\|\leq 1$ is introduced here to avoid the scale ambiguity issue. A good signal reconstruction can be easily obtained by the linear combinations of a few certain atoms.

Building on similar principles as dictionary learning, CDL replaces matrix multiplication with convolution operation for feature learning and signal reconstruction. Let $\dvec_k \in \mathbb{R}^M$ be the $k$-th atom in the dictionary $\{ \dvec_k \}_{k=1}^K$ with the total number of $K$ atoms, $\mathbf{s}_{n,k} \in \mathbb{R}^T$ be the corresponding sparse coefficient. The CDL is formulated as,
\begin{align}\label{eq:cdl}
    \begin{aligned}
        \min_{\bbD, \bbs_n} \sum_n^N \| \bbx_n - \ccalP_B \left\{\sum_{k=1}^{K} \dvec_k \ast \bbs_{n,k} \right\} \|_2^2 + \\ \lambda \sum_{k=1}^{K}\|\bbs_{n,k}\|_1,\ \ \ \  \|\dvec_k\|\leq 1,
    \end{aligned}
\end{align}

where $\ast$ denotes the convolution operator, and $\ccalP_B\{\cdot\}$ is a truncation operator pruning the signal of length $M+T-1$ to $T$ by removing the boundaries. For example, if $M$ is odd, $\ccalP_B\{\tilde \bbx\} := \bbB \tilde \bbx$, where $\bbB := \left[{\bf 0}_{T \times \frac{M-1}{2}}\ \bbI_{T \times T} \ {\bf 0}_{T \times \frac{M-1}{2}}\right]$. A remarkable difference is the atom dimension $M$ does not need to be equal to the input signal dimension, and the sparse coefficients dimension does not need to be equal to the atom dimension. This flexibility enables the learning of local features. Although the convolution operator is translation-variant, its ability to apply shared atoms across all positions allows CDL to detect recurring patterns regardless of their location, thereby enabling the learning of translation-invariant features.

\subsection{Discriminative DL \& Convolutional Dictionary Learning}
\label{sec:discr_DL_CDL}
The standard DL-based data reconstruction methods are not tailored for classification tasks. For a supervised setting, because the label for each piece of data is available, supposing the data have a total of $C$~classes, we can partition the data matrix $\bbX = [\bbX^{(1)}, ..., \bbX^{(C)}]$. In the same way, the dictionary can be formulated as $\bbD = [\bbD\c[1],\ldots,\bbD\c[C]] \in \mathbb{R}^{T \times K}$, where $K = K_1 + \ldots + K_C$ and $K_c$ is the number of atoms in $\bbD\c[c]$, and the sparse coefficients $\bbZ = [\bbZ\c[1],\ldots,\bbZ\c[C]] \in \mathbb{R}^{K \times N}$. For a supervised classification problem, Fisher Discriminative Dictionary Learning (FDDL) \cite{FDDL} is one of the most popular Discriminative DL methods, which is formulated as:
\begin{align}
    \min_{\bbD, \bbZ} r(\bbX, \bbD, \bbZ) + \lambda_1 \|\bbZ\|_1 +\lambda_2 f(\bbZ), \ \ \ \  \|\bbd_k\c[c]\|\leq 1
\end{align}
Here, the discriminative fidelity term is defined as: 
$r(\bbX, \bbD, \bbZ) = \sum_{c=1}^C \|\bbX^{(c)} - \bbD \bbZ^{(c)}\|_F^2 + \|\bbX^{(c)} -\bbD^{(c)} \bbZ^{(c)}_c\|_F^2 + \sum_{i \neq c} \| \bbD^{(i)}\bbZ^{(c)}_i\|^2_F$ , which contains the $\bbZ^{(c)}_c$ and $\bbZ^{(c)}_i$, namely the corresponding rows of sparse coefficients for label $c$ and labels except $c$. The fisher term $f(\bbZ)$ is defined as $f(\bbZ) = \sum_{c=1}^C \|\bbZ^{(c)}- \bbM^{(c)}\|_F - \|\bbM^{(c)}-\bbM\|_F + \|\bbZ\|^2_F$, where $\bbM^{(c)}$ and $\bbM$ are composed of the copies of the mean vectors of $\bbZ^{(c)}$ and $\bbZ$ respectively.

To solve the classification problem, a CDL framework is proposed in~\cite{GRK07}. Unlike the FDDL with discriminative fidelity term and fisher term, \cite{GRK07} specifies that each training example is postulated to be well reconstructed by the corresponding class atoms. The traning formulation is presented as

\begin{align} \label{convd}
    \begin{aligned}
        \min_{ \{\dvec_k\}, \bbs_{n,k}\c[c]} \sum_n^N \| \bbx_n - \sum_{c=1}^{C}\ccalP_B \left\{\sum_{k=1}^{K} \dvec_k\c[c] \ast \bbs_{n,k}\c[c] \right\} \|_2^2 + \\ \lambda \sum_{c=1}^{C}\sum_{k=1}^{K}\|\bbs_{n,k}\c[c]\|_1 . \ \ \ \ s.t. \ \|\dvec_k\c[c]\|\leq 1,
    \end{aligned}
\end{align}

As one can see that CDL formulation can be over-complicated with current notation. Let us define a set of notations and operations to simplify the formulation. Let $\mathcal{D}:= [\mathcal{D}\c[1], ..., \mathcal{D}\c[C] ] \in \mathbb{R}^{M \times K}$ with each per-class dictionaries $\mathcal{D}\c[c] \in \mathbb{R}^{M \times K_c}$ for class index $c \in \{1,2,\ldots,C\}$, for each example the sparse coefficients be $\bbS_n := [\bbS_n\c[1], ..., \bbS_n\c[C] ] \in \mathbb{R}^{T \times K}$ with per-class sparse coefficients $\bbS_n\c[c] \in \mathbb{R}^{T \times K_c}$, the convolution operator $*$ for the matrices be $ \bbH * \bbQ := \sum_{l=1}^{L}\ccalP_B\{\bbh_l * \bbq_l \}, \bbH = [\bbh_1,..., \bbh_L] \in \mathbb{R}^{M\times L}, \bbQ =[\bbq_1, ..., \bbq_L] \in \mathbb{R}^{T \times L}$. So the formulation (\ref{convd}) can be re-written as
\begin{align}\label{convd_compact}
    \min_{ \mathcal{D}, \bbS_n} \sum_n^N \| \bbx_n -  \mathcal{D} \ast  \bbS_n \|_2^2 + \lambda \| \bbS_n\|_1 \ \  \ \ s.t. \ \|\dvec_k\c[c]\|\leq 1.
\end{align}

\subsection{MIML CDL with background handling}
\label{sec:wscdl}

Separating the common features from the overall learned features will facilitates more effective extraction of the discriminative features, thereby improving classification performance. To model the background as well as the discriminative components in the mixture, we adopt a similar idea to \cite{VuM17}, learning both the shared dictionary $\mathcal{D}\c[0] \in \mathbb{R}^{M \times K_0}$ and the discriminative dictionary $\mathcal{D}$. The combined dictionary is defined as $ \mathcal{\bar D} := [\mathcal{D}\c[0], \mathcal{D}] \in \mathbb{R}^{M \times \bar K}$ with $\bar K = K_0 + K$. With $\bar \bbS_n := [\bbS_n\c[0],\bbS_n] \in \mathbb{R}^{T \times \bar K}$, the formulation (\ref{convd_compact}) with shared dictionary can be formulated as:
\begin{align}
    \begin{aligned}
        \min_{ \mathcal{\bar D}, \bar \bbS_n} \sum_n^N \| \bbx_n - \mathcal{\bar D} \ast \bar \bbS_n \|_2^2 + \lambda \|\bar \bbS_n\|_1 + \mu \|\mathcal{D}\c[0]\|_*, \\ \ \ \ \ s.t. \ \|\dvec_k\c[c]\|\leq 1, c= 0,...,C
    \end{aligned}
\end{align}
where $\|\bar \bbS_n\|_1$ is the sum of the absolute values of all entries in $\bar \bbS_n$, which promotes sparsity, and $\|\mathcal{D}\c[0]\|_*$ is the nuclear norm (sum of singular values) of $\mathcal{D}\c[0]$, which promotes low rank. Low rankness is necessary to prevent the common dictionary from absorbing discriminant atoms. Otherwise, $\mathcal{D}\c[0]$ could contain all the discriminative information, which means $\| \bbx_n - \mathcal{D}\c[0] \ast \bbS_n\c[0] \|_2^2$ could be very small without the regularizer $\|\mathcal{D}\c[0] \|_*$.

However, to address the problem with {\em MIML} settings, in the sense that $\{\bbx_n\}_1^N$ cannot be partitioned into $C$~parts, auxiliary label vectors are needed. Specifically, the supervision can be provided in the form of a $C$-dimensional binary vector $\bby_n \in \{0,1\}^C$, where the $c$-th element of $\bby_n$ is $1$ if a class-$c$ signal is present in $\bbx_n$, and $0$ otherwise. Note that $\bby_n$ may be an all-one vector or all-zero vector, which means that $\bbx_n$ contains all the classes or only background noise, respectively.

In the papers~\cite{FDDL,VuM17}, they have demonstrated the importance of using discriminative fidelity, instead of just using per-class reconstruction as in formulation (\ref{convd_compact}). Furthermore, to fit the MIML settings, the discriminative fidelity should have the capability to pick multiple class reconstructions. We now define our fidelity term $\ccalF(\bbx_n,\bby_n,\mathcal{\bar D}, \bar \bbS_n)$ as:
\begin{align}
    \begin{aligned} \label{fxd}
        \ccalF(\bbx_n,\bby_n,\mathcal{\bar D}, \bar \bbS_n) := \left\|\bbx_n - \mathcal{\bar D} \ast \bar \bbS_n \right\|^2        \\
        + \left\|\bbx_n - \sum_{c:\, y_n\c[c]=1} \mathcal{D}\c[c] \ast \bbS_n\c[c] - \mathcal{D}\c[0] \ast \bbS_n\c[0]  \right\|^2 \\
        + \sum_{c':\, y_n\c[c'] = 0} \|\mathcal{D}\c[c'] \ast \bbS_n\c[c'] \|^2,
    \end{aligned}
\end{align}
where the first term is the reconstruction error for $\bbx_n$ using the entire dictionary $\mathcal{\bar D}$, the second term is the reconstruction error using the common dictionary for the background and the per-class dictionaries only corresponding to the signal labels that are present in $\bbx_n$, and finally the third term is the residual due to the per-class dictionaries for the signal classes that are {\em not} present in $\bbx_n$.

Compared with~\cite{FDDL,VuM17}, instead of using the Fisher term for single label classification, we need a label penalty term to match the MIML settings. To predict the class labels $\bby_n$, one solution is to project the discriminative sparse signal $\bbS_n$ to $C$ dimensions and perform adaptive ``pooling" over time. Specifically, we define weight vectors $\bbw\c[c] \in \mathbb{R}^{K_c}$, for $c=1,2,\ldots,C$ and the bias term $\bbb \in \mathbb{R}^C$. So the projection $\bbW := \mathrm{bdiag}\{\bbw\c[1],\ldots,\bbw\c[C]\} \in \mathbb{R}^{K \times C}$ has a block diagonal structure. Then the predicted label $\hat \bby_n$ is given by:
\begin{align}
    \hat \bby_n =  \frac{1}{1 + e^{ \ccalP_{P}\{ \bbS_n \bbW \}^\top + \bbb }}  \in \mathbb{R}^C
\end{align}
where $\cdot^\top$ denotes transposition and $\ccalP_P$ is a pooling operator, such as the average pooling $\ccalP_{P_{avg}}\{\tilde \bbY\} = T^{-1} {\bf 1}_{1 \times T} \tilde \bbY$ and $\ccalP_{P_{max}}\{\tilde \bbY\} = [\max\{\tilde \bby\c[1]\},\ldots,\max\{\tilde \bby\c[C]\}]^\top$, where $\tilde \bby\c[c]$ is the $c$-th row of $\tilde \bbY$. The label penalty function $\ccalL(\hat \bby,\bby)$ is used to assess the discrepancy between the true label $\bby = [y\c[1],\ldots,y\c[C]]^\top$ and the predicted label $\hat \bby = [\hat y\c[1],\ldots,\hat y\c[C]]^\top$ can be chosen, for instance, as the hinge loss:
\begin{align}
    \ccalL_{hg}(\hat \bby,\bby) := \sum_{c=1}^C \max\{0, 1-2(\hat y\c[c] -1)(2 y\c[c]-1)\}
\end{align}
or the cross-entropy loss:
\begin{align}
    \ccalL_{cr}(\hat \bby,\bby) := -\sum_{c=1}^C [y\c[c]\log \hat y\c[c] + (1-y\c[c])\log(1-\hat y\c[c])].
\end{align}

Then, the overall optimization problem for training $\mathcal{\bar D}$, $\bar \bbS_n$, $\bbW$ and $\bbb$, with given $N$ examples $\{\bbx_n,\bby_n\}_{n=1}^N$ can be formulated as:
\begin{align}
    \begin{aligned}
        \min_{\mathcal{\bar D},\bbW, \bbb} \sum_{n=1}^N \min_{\bar \bbS_n} \Big\{\ccalF(\bbx_n,\bby_n,\mathcal{\bar D}, \bar \bbS_n) + \lambda \|\bar \bbS_n \|_1 + \eta \ccalL(\hat \bby_n,\bby_n) \Big\} \\
        + \mu \|\mathcal{D}\c[0] \|_*, \quad \text{s.t. } \|\dvec_k\c[c]\|\leq 1, c= 0,...,C \label{costfun}
    \end{aligned}
\end{align}
This formulation is based on $\bbx_n$ being a $T$-dimensional vector. If each of the input data is a matrix or a high-dimensional tensor, more discussion will be presented in Sec.~\ref{sec:ext}.
\section{Algorithm Derivation}
\label{sec:method}
\subsection{BPG-M setup}
\label{sec:bpgm}
In this section, we adopt the BPG-M as the method mentioned in~\cite{ChunF18} to solve the proposed formulation (\ref{costfun}). The BPG-M is trying to solve the minimization problem
\begin{align} \label{general problem}
	\min F(\bbx_1, ..., \bbx_k) = f(\bbx_1, ..., \bbx_k) + \sum_{k}^{K}g_k(\bbx_k),
\end{align}
where $f$ is assumed to be continuously differentiable, but functions $g_k$ with $k =\{ 1, ..., K\}$ are not necessarily differentiable. There is no assumption that either $f$ or $g_k$ with $k =\{ 1, ..., K\}$ needs to be convex. Suppose $\nabla f$,
the gradient of $f$, is M-Lipschitz continuous, which means $\nabla f$ satisfies
\begin{align}
	\| \nabla f(\bbx) - \nabla f(\bby) \|_{\bbM^{-1}} \leq \|\bbx-\bby\|_\bbM,\ \bbx,\bby \in \mathbb{R}^n, \nonumber \\
	\text{where } \|\bbx\|^2_\bbM := \bbx^T \bbM \bbx.
\end{align}
The $f$ has a quadratic majorization via M-Lipschitz continuous gradients as
\begin{align}
	f(\bbx) \leq f(\bby) + <\nabla f(\bby), \bbx-\bby> + \frac{1}{2}\|\bbx- \bby\|^2_\bbM
\end{align}
How to choose and calculate the majorization matrix $\bbM$ will be discussed in the later section \ref{sec:bpgm_app}. Since there exists an upper bound of $f$, in order to solve the minimization problem (\ref{general problem}), iteratively lowering the upper bound helps to minimize the overall function. The proximal operator with majorization matrix is applied to achieve this goal, with given $\bbM$, and the proximal operator is
\begin{align}
	\text{Prox}_g(\bby; \bbM) = \arg\min_\bbx \frac{1}{2} \|\bbx - \bby\|_\bbM^2 + g(\bbx).
\end{align}
Let us define, in the $i$-th update, for the $k$-th block, the cost function can be written as
\begin{align}
	f_k\c[i+1](\bbx_k) = f(\bbx_1\c[i+1] ... , \bbx_{k-1}\c[i+1],\bbx_k, \bbx_{k+1}\c[i] ..., \bbx_K\c[i]).
\end{align}
The update for $\bbx_k\c[i+1]$ can be calculated through the following update
\begin{align}
	\bbx_k\c[i+1] & = \arg\min_{\bbx_k\c[i]} <\nabla f_k\c[i+1](\tilde \bbx_k\c[i]),\bbx_k\c[i] - \tilde \bbx_k\c[i] >  \nonumber \\
	              & + \frac{1}{2} \|\bbx_k\c[i] - \tilde \bbx_k\c[i]\|_{\bbM_k^{(i)}}^2 +g_k(\bbx_k\c[i])\nonumber              \\
	              & = \text{Prox}_g( \tilde \bbx_k^{(i)} - {\bbM_k^{(i)}}^{-1} \nabla f(\tilde \bbx_k^{(i)}) ; \bbM_k^{(i)})
\end{align}
where $\tilde \bbx_k^{(i)} = \bbx_k^{(i)} + \bbM_{W_k}^{(i)}(\bbx_k^{(i)} - \bbx_{k}^{(i-1)})$ with the step size $\bbM_{W_k}^{(i)} = \delta {\bbM_k^{(i)}}^{-1/2}{\bbM_{k}^{(i-1)}}^{1/2}, \ \  0<\delta<1 $. Then the general BPG-M algorithm can be written in table \ref{alg:bpgm}.

There are several benefits of using BPG-M to solve the problem, as claimed in Chun and Fessler's papers \cite{ChunF18, chun19}. One of them is, unlike FISTA \cite{FISTA},  the function $g$ or $f$ does not to be convex. Another attractive attribute is the guarantee of convergence, though it may converge to a local minimum. In addition, it claims to be faster than other comparative methods\cite{ChunF18}, in terms of time complexity and actual calculation time.


\begin{algorithm}[t]
\caption{General BPG-M Algorithm}
\begin{algorithmic}[1]
\STATE Initialize $i=1$
\WHILE{not converged}
    \FOR{$k = 1$ to $K$}
        \STATE Compute $\bbM_k^{(i)}$
        \STATE $\bbM_{W_k}^{(i)} \gets \delta {\bbM_k^{(i)}}^{-1/2} {\bbM_{k}^{(i-1)}}^{1/2}$ \hfill ($0<\delta<1$)
        \STATE $\tilde \bbx_k^{(i)} \gets \bbx_k^{(i)} + \bbM_{W_k}^{(i)}(\bbx_k^{(i)} - \bbx_{k}^{(i-1)})$
        \STATE $\nu \gets \tilde \bbx_k^{(i)} - {\bbM_k^{(i)}}^{-1} \nabla f(\tilde \bbx_k^{(i)})$
        \STATE $\bbx_k^{(i+1)} \gets \arg\min_{\bbx} \frac{1}{2} \|\bbx - \nu\|_{\bbM_k^{(i)}}^2 + g(\bbx)$
    \ENDFOR
    \STATE $i \gets i + 1$
\ENDWHILE
\end{algorithmic}
\label{alg:bpgm}
\end{algorithm}

\subsection{Solving the problem with BPG-M}
\label{sec:bpgm_app}

Firstly, in order to apply BPG-M, the convolution operation needs to be reformulated as matrix multiplication by representing it as a Toeplitz matrix. Then, each of $\dvec_k^{(0)}$, $\dvec_k^{(c)}$, $\bbs_{n,k}^{(0)}$, $\bbs_{n,k}^{(c)}$, and $\bbw_k^{(c)}$ is alternately solved by computing the majorizer function and majorizer matrix for each variable. Finally, the algorithm for solving (\ref{costfun}) using BPG-M is presented. Here, we use the cross-entropy loss, as it provides the probability of the label for a given class. Average pooling is employed as the pooling operator.. For simplicity, the update of the bias term $\bbb$ is included in the variable $\bbW$; more details can be found in Section~\ref{sec:updataW}.

\subsubsection{Shared dictionary update}
\label{sec:updataD0}
when updating $\mathcal{D}^{(0)}$ with all other variables fixed, the cost function is:
\begin{equation} \label{cost_d0}
	\begin{aligned}
		\min_{\dvec_k^{(0)}} & \sum_{n=1}^{N} \left\|\bbx_n - \mathcal{\bar D} \ast \bar \bbS_n \right\|^2 + \mu \|\mathcal{D}^{(0)} \|_*\\
		& + \left\|\bbx_n - \sum_{c:\, y_{n,c}=1} \mathcal{D}^{(c)} \ast \bbS_n^{(c)} - \mathcal{D}^{(0)} \ast \bbS_n^{(0)}  \right\|^2 , \\
		& \text{s.t.} \ \|\dvec_k^{(0)}\|\leq 1,
	\end{aligned}
\end{equation}
which could be rewritten as matrix multiplication using the truncated Toeplitz matrix as $\bbT_{s_{n,k}\c[0]} \in \mathbb{R}^{T \times M}$, where $\bbT_{s_{n,k}\c[0]}$ can be calculated as
\begin{align}\label{Tsnk0}
	\begin{aligned}
		\bbT_{s_{n,k}\c[0]} = {\ccalP_B} \Big\{ \Big[ & \Big[{\bf 0}_{1 \times \frac{M-1}{2}}, \bbs_{n,k}^{(0) \top } \Big]^{\top},                      \\
		& \Big[{\bf 0}_{1 \times \frac{M-3}{2}}, \bbs_{n,k}^{(0) \top}, {\bf 0}_{1 \times 1} \Big]^{\top}, \\
		& \ldots, \Big[\bbs_{n,k}^{(0) \top } ,{\bf 0}_{1 \times \frac{M-1}{2}}\Big]^{\top} \Big] \Big\}.
	\end{aligned}
\end{align}
The formulation (\ref{cost_d0}) could be rewritten as
\begin{equation} \label{updateD0}
	\begin{aligned}
		\min_{\dvec_k^{(0)}} & \sum_{n=1}^{N} \frac{1}{2} \|2\bbx_n - \alpha_n - \beta_n - 2\bbT_{s_{n,k}^{(0)}} \dvec_k^{(0)} \|_2^2 + \mu \|\mathcal{D}^{(0)}\|_*, \\
		& \text{s.t.} \ \|\dvec_k^{(0)}\|\leq 1,
	\end{aligned}
\end{equation}
where  $ \alpha_n$ and $\beta_n$ are respectively defined as $ \alpha_n =   \mathcal{D} \ast  \bbS_n + \mathcal{D}^{(0)'}  \ast  \bbS^{(0)'}  $ and $ \beta_n =   \sum_{c=1}^{C} y_n\c[c] \mathcal{D}\c[c] \ast  \bbS\c[c]_n  + \mathcal{D}^{(0)'}  \ast  \bbS^{(0)'}$. Here $\mathcal{D}^{(0)'}  \ast  \bbS^{(0)'} $ denotes $\mathcal{D}^{(0)'}  \ast  \bbS^{(0)'}$ without ${\ccalP_B} \left\{ \dvec_k^{(0)}  \ast  \bbs_{n,k}^{(0)} \right\}$.

Algorithm~\ref{alg:bpgm} can be applied to solve this problem (\ref{updateD0}) with a fixed majorization matrix $\bbM_{\dvec_k\c[0]} = \text{diag}\{\sum_{n=1}^{N} 4|\bbT_{s_{n,k}\c[0]}|^\top |\bbT_{s_{n,k}\c[0]}|{\bf 1}_{M \times 1} \}$~\cite{chun19}, resulting in a step size of simply $\delta \bbI$. With $\nu_{\dvec_k\c[0]} := \tilde\dvec_k\c[0] - \bbM_{\dvec_k\c[0]}^{-1} \sum_n 2 \bbT_{s_{n,k}\c[0]}^{\top} (2 \bbT_{s_{n,k}\c[0]} \tilde\dvec_k\c[0] - 2\bbx_n + \alpha_n + \beta_n)$ and $\tilde\dvec_k\c[0]$ calculated in line (5) of Algorithm~\ref{alg:bpgm} using $\dvec_k\c[0]$ from the previous iteration, the sub-problem for the BPG-M algorithm is to solve the proximal operator, which is

\begin{align} \label{dk0_sub}
	\begin{aligned}
		\dvec_k\c[0] = \arg\min_{\dvec_k\c[0]} \frac{1}{2} \| \dvec_k\c[0] - \nu_{\dvec_k\c[0]} \|^2_{\bbM_{\dvec_k\c[0]}} + N\mu \|\mathcal{D}\c[0]\|_*, \ \ \\ \ \ s.t. \ \|\dvec_k\c[0]\|\leq 1.
	\end{aligned}
\end{align}
To solve problem (\ref{dk0_sub}), here we applied ADMM \cite{BCE10} and problem (\ref{dk0_sub}) can be reformulated as
\begin{align}
	\begin{aligned}
		\min_{\dvec_k\c[0]} \frac{1}{2} \| \dvec_k\c[0] - \nu_{\dvec_k\c[0]} \|^2_{\bbM_{\dvec_k\c[0]}} + N\mu \|\bbO\|_*, \ \ \\ \ \ s.t. \ \|\dvec_k\c[0]\|\leq 1, \mathcal{D}\c[0] = \bbO,
	\end{aligned}
\end{align}
\begin{align} \label{admm}
	\begin{aligned}
		\min_{\dvec_k\c[0]} \frac{1}{2} \| \dvec_k\c[0] - \nu_{\dvec_k\c[0]} \|^2_{\bbM_{\dvec_k\c[0]}} + N\mu \|\bbO\|_* + \\ \text{tr}(\bbY^\top(\bbO-\mathcal{D}\c[0])) + \frac{\rho}{2}\| \bbO-\mathcal{D}\c[0] \|^2_F, \ \ \\ \ \ s.t. \ \|\dvec_k\|\leq 1.
	\end{aligned}
\end{align}
Now one can solve this problem (\ref{admm}) by the following steps, iteratively until converged: 1. update $\dvec_k\c[0]$ by solving the problem $\min_{\dvec_k\c[0]} \frac{1}{2} \| \dvec_k\c[0] - \nu \|^2_{\bbM_{\dvec_k\c[0]}} + \frac{\rho}{2} \|\bbz_k-\dvec_k\c[0]+ \frac{1}{\rho}\bby_k\|_2^2$ with the solution for the Quadratically Constrained Quadratic Program (QCQP) \cite{BV04} 
2. update $\bbO$, through $N\mu \|\bbO\|_* + \frac{\rho}{2}\|\bbO - \mathcal{D}\c[0]+\frac{1}{\rho}\bbY\|_F^2$ solved by a singular value thresholding SVT \cite{CCS10}
3. update $\bbY$, simply by $\bbY = \bbY + \rho (\bbO-\mathcal{D}\c[0])$, where $\rho > 1$.

\begin{algorithm}[t]
	\caption{Algorithm for shared dictionary update}
	\begin{algorithmic}[1]
		\STATE Initialize $\mathcal{\bar D}, \bar \bbS_n, \bbW, \bbb$ with random values
		\WHILE{not converge}
		\STATE /* Update $\mathcal{D}\c[0]$ */
		\FOR{all $\dvec_k\c[0]$ in $\mathcal{D}\c[0]$}
		\STATE Calculate truncated Toeplitz matrix $\bbT_{s_{n,k}\c[0]}$ by (\ref{Tsnk0})
		\STATE Calculate majorization matrix \\
		$\bbM_{\dvec_k\c[0]} = diag\{\sum_{n=1}^{N} 4|\bbT_{s_{n,k}\c[0]}|^\top |\bbT_{s_{n,k}\c[0]}|{\bf 1}_{M \times 1} \}$
		\STATE Step size is $\delta \bbI$
		\ENDFOR
		\STATE /* Check convergence */
		\IF{overall loss changing rate $\leq \epsilon$}
		\STATE break
		\ENDIF
		\ENDWHILE
	\end{algorithmic}
	\label{alg:shared_dict_update}
\end{algorithm}

\subsubsection{Discriminative dictionary update}
\label{sec:updataD}
The cost function for each of the $\dvec_k\c[c]$ in $\mathcal{D}$, with all other variables fixed, is the following
\begin{equation} \label{updateD}
	\begin{aligned}
		\min_{\dvec_k\c[c]} \quad & \sum_{n=1}^{N} \left\|\bbx_n - \mathcal{\bar D} \ast \bar \bbS_n \right\|^2 + \sum_{c':\, y_n\c[c'] = 0} \|\mathcal{D}\c[c'] \ast \bbS_n\c[c'] \|^2 \\
		& + \left\|\bbx_n - \sum_{c:\, y_n\c[c]=1} \mathcal{D}\c[c] \ast \bbS_n\c[c] - \mathcal{D}\c[0] \ast \bbS_n\c[0]  \right\|^2 \\
		& \text{s.t.} \ \|\dvec_k\c[c]\|\leq 1,
	\end{aligned}
\end{equation}
The formulation (\ref{updateD}) is equivalent to
\begin{equation}
	\begin{aligned}
		\min_{\dvec_k\c[c]} & \sum_{n=1}^{N} \left\|\bbx_n - \mathcal{\bar D} \ast \bar \bbS_n \right\|^2                \\
		& + \left\|\bbx_n - \sum_{c} y_n\c[c]\mathcal{D}\c[c] \ast \bbS_n\c[c] - \mathcal{D}\c[0] \ast \bbS_n\c[0]  \right\|^2 \\
		& + \sum_{c} (1-y_n\c[c])\|\mathcal{D}\c[c] \ast \bbS_n\c[c] \|^2,                      \\
		& \text{s.t.} \ \|\dvec_k\c[c]\|\leq 1,
	\end{aligned}
\end{equation}
which could be further reduced to
\begin{equation} \label{cost_dkc}
	\begin{aligned}
		\min_{\dvec_k\c[c]} \sum_{n=1}^{N} \frac{1}{2} \| \bbT_{s_{n,k}\c[c]} \dvec_k\c[c] -\gamma_n \|_2^2, \ \  \ \ s.t. \ \|\dvec_k\c[c]\|\leq 1
	\end{aligned}
\end{equation}
where the Toeplitz matrix of $\bbs_{n,k}\c[c]$ is defined as
\begin{equation}\label{Tsnkc}
	\begin{aligned}
		\bbT_{s_{n,k}\c[c]} := {\ccalP_B} \Big\{ \Big[ & \Big[{\bf 0}_{1 \times \frac{M-1}{2}}, \bbs_{n,k}^{(c) \top } \Big]^{\top},                      \\
		& \Big[{\bf 0}_{1 \times \frac{M-3}{2}}, \bbs_{n,k}^{(c) \top}, {\bf 0}_{1 \times 1} \Big]^{\top}, \\
		& \ldots, \Big[\bbs_{n,k}^{(c) \top } ,{\bf 0}_{1 \times \frac{M-1}{2}}\Big]^{\top} \Big] \Big\}.
\end{aligned}
\end{equation}
and $\gamma_n$ can be calculated as
\begin{equation}
	\begin{aligned}\label{gamma}
		\gamma_n := \frac{1}{2}( & \bbx_n - \mathcal{\bar D}' \ast \bar \bbS_n'\\
		& + y_n\c[c](\bbx_n - \mathcal{D}\c[0]*\bbS_n\c[0] - y_n\c[c]{\mathcal{D}\c[c]}' \ast {\bbS_n\c[c]}' \\
		& -  \sum_{c' \neq c} y_n\c[c']\mathcal{D}\c[c'] \ast \bbS_n\c[c'])   \\
		& - (1-y_n\c[c])((1-y_n\c[c]){\mathcal{D}\c[c]}' \ast {\bbS_n\c[c]}'  \\
		& + \sum_{c' \neq c} (1-y_n\c[c'])\mathcal{D}\c[c'] \ast \bbS_n\c[c'])).
	\end{aligned}
\end{equation}
In the formulation (\ref{gamma}), $\mathcal{\bar D}' \ast \bar \bbS_n'$ denotes the $\mathcal{\bar D} \ast \bar \bbS_n$ without the term $\ccalP_B \{\dvec_k\c[c] * \bbs_{n,k}\c[c]\}$. Similarly, ${\mathcal{D}\c[c]}' \ast {\bbS_n\c[c]}'$ means ${\mathcal{D}\c[c]} \ast {\bbS_n\c[c]}$without $\ccalP_B \{\dvec_k\c[c] * \bbs_{n,k}\c[c]\}$.
To follow the procedure described in the Algorithm \ref{alg:bpgm}, the majorization matrix can be simply $\bbM_{\dvec_k\c[c]} = \text{diag}\{\sum_{n=1}^{N} |\bbT_{s_{n,k}\c[c]}|^\top |\bbT_{s_{n,k}\c[c]}|{\bf 1}_{M \times 1} \} $ and step size just chosen as $\delta \bbI$. Let us define $\nu_{\dvec_k\c[c]} := \tilde\dvec_k\c[c] - \bbM_{\dvec_k\c[c]}^{-1} \sum_n \bbT_{s_{n,k}\c[c]}^{\top} ( \bbT_{s_{n,k}\c[c]} \tilde\dvec_k\c[c] - \gamma_n)$ and $\tilde\dvec_k\c[c]$ is still calculated by line (5) in Algorithm \ref{alg:bpgm}. The last step is to solve the problem
\begin{align} \label{dkc_sub}
	\dvec_k\c[c] = \arg\min_{\dvec_k\c[c]} \frac{1}{2} \| \dvec_k\c[c] - \nu_{\dvec_k\c[c]} \|^2_{\bbM_{\dvec_k\c[c]}} , \ \  \ \ s.t. \ \|\dvec_k\c[c]\|\leq 1.
\end{align}
which is still a QCQP problem that can be solved by accelerated Newton's method \cite{ChunF18}.

\subsubsection{Common sparse coefficients update}
\label{sec:updataS0}
In order to update the common sparse coefficients, the cost function for $\bbs_{n,k}\c[0]$ can be reformulated as
\begin{equation}
	\begin{aligned}
		\min_{\bbs_{n,k}\c[0]} & \left\|\bbx_n - \mathcal{\bar D} \ast \bar \bbS_n \right\|^2 \\
		& + \left\|\bbx_n - \sum_{c:\, y_n\c[c]=1} \mathcal{D}\c[c] \ast \bbS_n\c[c] - \mathcal{D}\c[0] \ast \bbS_n\c[0]  \right\|^2 \\
		& + \lambda \|\bbs_{n,k}\c[0] \|_1,
	\end{aligned}
\end{equation}
\begin{equation}\label{updatas0}
	\begin{aligned}
		\min_{\bbs_{n,k}\c[0]} \frac{1}{2} \|2\bbx_n - \alpha_n - \beta_n
		- 2\bbT_{d_k\c[0]} \bbs_{n,k}\c[0] \|_2^2+ \lambda \|\bbs_{n,k}\c[0] \|_1.
	\end{aligned}
\end{equation}
where
\begin{equation}\label{Td0}
	\begin{aligned}
		\bbT_{d_k\c[0]} = {\ccalP_B} \Big\{ \Big[ & \Big[{\bf 0}_{1 \times \frac{T-1}{2}}, \dvec_k^{(0) \top } \Big]^{\top},     \\
		& \Big[{\bf 0}_{1 \times \frac{T-3}{2}}, \dvec_k^{(0) \top}, {\bf 0}_{1 \times 1} \Big]^{\top},   \\
		& \ldots, \Big[\dvec_k^{(0) \top } ,{\bf 0}_{1 \times \frac{T-1}{2}}\Big]^{\top} \Big] \Big\} \in \mathbb{R}^{T \times T}
	\end{aligned}
\end{equation}
is the truncated Toeplitz matrix of $\dvec_k\c[0]$. One noticeable advantage of updating $\bbs_{n,k}\c[0]$ is it could be parallelly computed for each trainig example index $n$. Applying BPG-M to solve problem (\ref{updatas0}), the majorization matrix could be chosen as $\bbM_{\bbS\c[0]} = \text{diag}\{ 4|\bbT_{d_k\c[0]}|^\top |\bbT_{d_k\c[0]}|{\bf 1}_{T \times 1}\}$, and the step size is still a constant $\delta \bbI$. With $\nu_{\bbs_{n,k}\c[0]}$ defined as $\nu_{\bbs_{n,k}\c[0]}:= \tilde\bbs_{n,k}\c[0] - {\bbM_{\bbS^{(0)}}}^{-1} 2\bbT_{d_k\c[0]}^{\top}(2\bbT_{d_k\c[0]} \tilde\bbs_{n,k}\c[0] - 2\bbx_n + \alpha_n + \beta_n)$ and again $\tilde\bbs_{n,k}\c[0]$ calculated at line 5 in Algorithm \ref{alg:bpgm}. The last step of the BPG-M will solve the following problem
\begin{align}
	\min_{\bbs_{n,k}\c[0]} \frac{1}{2} \|\bbs_{n,k}\c[0] - \nu_{\bbs_{n,k}\c[0]} \|_{\bbM_{\bbS\c[0]}}^2+ \lambda \|\bbs_{n,k}\c[0] \|_1.
\end{align}
which can be solved by a threshold operator $\Gamma(a, b) := \text{sign}(a) \max(|a|-b, 0)$, where $\max(|a|-b, 0)$ means the maximum value between $|a|-b$ and $0$. The solution is simply $\Gamma(\nu_{\bbs_{n,k}\c[0]}(i), \lambda {\bbM_{\bbS^{(0)}}}^{-1}(i,i))$, where $\nu_{\bbs_{n,k}\c[0]}(i)$ is the $i$-th element of $\nu_{\bbs_{n,k}\c[0]}$ and ${\bbM_{\bbS^{(0)}}}^{-1}(i,i)$ is the element on the $i$-th row, $i$-th column of the inversed matrix of $\bbM_{\bbS\c[0]}$.

\subsubsection{Discriminative sparse coefficients update}
\label{sec:updataS}
Similar to the common sparse coefficients update, the update for each $\bbs_{n,k}\c[c]$ can be done in a parallel manner. Let's define $\bbp := T^{-1} {\bf 1}_{T \times 1}$ and $ \hat \bbs_n\c[c]$ can be written as $\hat \bbs_n\c[c] = [ \bbp^\top \bbs_n\c[c], 1 ]^{\top} $.
Let the truncated Toeplitz matrix of $\dvec_k\c[c]$ be defined as
\begin{align}\label{Tdck}
	\begin{aligned}
		\bbT_{d_k\c[c]} = {\ccalP_B} \Big\{ \Big[ & \Big[{\bf 0}_{1 \times \frac{T-1}{2}}, \dvec_k^{(c) \top } \Big]^{\top},\\
		& \Big[{\bf 0}_{1 \times \frac{T-3}{2}}, \dvec_k^{(c) \top}, {\bf 0}_{1 \times 1} \Big]^{\top},   \\
		& \ldots, \Big[\dvec_k^{(c) \top } ,{\bf 0}_{1 \times \frac{T-1}{2}}\Big]^{\top} \Big] \Big\} \in \mathbb{R}^{T \times T}
	\end{aligned}
\end{align}
To update $\bbs_{n,k}\c[c]$, we solve the following minimization problem:
\begin{align}\label{updates}
	\begin{aligned}
		\min_{\bbs_{n,k}\c[c]} & \frac{1}{2} \| \bbT_{d_k\c[c]} \bbs_{n,k}\c[c] -\gamma_n \|_2^2 + \lambda \|\bbs_{n,k}\c[c] \|_1 \\
			& + \eta (1-y_n\c[c])(\bbp^\top \bbs_{n,k}\c[c] w_k\c[c])        \\
			& + \eta \log (1+ e^{({\hat\bbs_n}^{(c)\top} \hat \bbw\c[c])}).
	\end{aligned}
\end{align}
The Hessian matrix for $\bbs_{n,k}\c[c]$ is given by
\begin{align}
	\bbH_{\bbs_{n,k}\c[c]} = 2\bbT_{d_k\c[c]}^\top \bbT_{d_k\c[c]} + \eta \frac{e^{ {\hat \bbs_n}^{(c)\top} \hat \bbw\c[c]}  }{(1+e^{ {\hat \bbs_n}^{(c)\top} \hat \bbw\c[c]})^2} (w_k\c[c])^2 \bbp \bbp^\top,
\end{align}
where $w_k\c[c]$ is the $k$-th element of vector $\bbw\c[c]$ and is a scalar. The majorization matrix is chosen as
$\bbM_{\bbS\c[c]} = \text{diag}(4\bbT_{d_k\c[c]}^\top \bbT_{d_k\c[c]} + \eta \frac{1}{4} (w_k\c[c])^2 \bbp \bbp^\top)$.
In the last step of the BPG-M algorithm, we solve the problem (\ref{updats_last}) using the threshold operator $\Gamma$.
\begin{align} \label{updats_last}
	\min_{\bbs_{n,k}\c[c]} \frac{1}{2} \|\bbs_{n,k}\c[c] - \nu_{\bbs_{n,k}\c[c]} \|_{\bbM_{\bbS\c[c]}}^2+ \lambda \|\bbs_{n,k}\c[c] \|_1.
\end{align}
where $\nu_{\bbs_{n,k}\c[c]}$ is calculated as follows:
\begin{align}
	\label{nusnkc}
	\begin{aligned}
		\nu_{\bbs_{n,k}\c[c]} = \tilde \bbs_{n,k}\c[c] - \bbM_{\bbS\c[c]} ^{-1}( & 2\bbT_{d_k}^{(c)\top} (\bbT_{d_k\c[c]}\tilde \bbs_{n,k}\c[c] - \gamma_n)    \\
		& + \eta (1-y_n\c[c]) w_k\c[c] \bbp              \\
		& + \eta (1+ e^{\hat{\tilde\bbs}_n^{(c)\top} \hat\bbw\c[c]})^{-1}             \\
		& \quad \times e^{ \hat{\tilde\bbs}_n^{(c)\top} \hat\bbw\c[c]} w_k\c[c] \bbp)
\end{aligned}
\end{align}
where $\hat {\tilde \bbs}_n\c[c] = [ \bbp^\top \tilde\bbs_n\c[c], 1 ]^{\top}$ and $\tilde \bbs_{n,k}\c[c]$ is obtained from the previous steps during the BPG-M calculation. 

\subsubsection{Projection update}
\label{sec:updataW}
The loss function is chosen as the cross-entropy loss as mentioned before. With all other variables fixed, the loss function for $\bbW$ can be written as
\begin{align}
	\begin{aligned}
		\min_{\bbW, \bbb} \sum_{n=1}^{N} & ({\bf{1}} -\bby_n)^{\top}( \ccalP_{P}\{ \bbS_n \bbW \} + \bbb)    \\
		& + {\bf{1}}^{\top}\log(1+ e^{ \ccalP_{P}\{ \bbS_n \bbW \} + \bbb})
	\end{aligned}
\end{align}
To simplify the optimization, the bias term $\bbb$ is incorporated into $\bbW$. This bias term is crucial for preventing the predicted label probability from being stuck at 0.5 due to zeros in the sparse coefficients. We define $\hat \bbw\c[c] := [\bbw^{[c]\top},  b\c[c]]^{\top} \in \mathbb{R}^{K+1}$ and $ \hat \bbs_n\c[c] := [\ccalP_{P}\{ \bbs_n\c[c] \}, 1]^{\top} \in \mathbb{R}^{K+1} $. Consequently, the loss function for each $\hat \bbw\c[c]$ can be expressed as
\begin{align}
	\min_{\hat\bbw\c[c]} \sum_{n=1}^{N}  -(1-y_n\c[c])({\hat\bbs_n}^{(c)\top} \hat \bbw\c[c]) + \log(1+ e^{ {\hat \bbs_n}^{(c)\top} \hat \bbw\c[c]}).
\end{align}
Using the average pooling operator, the Hessian matrix for $\hat \bbw\c[c]$ is given by
\begin{align}
	\bbH_{\hat\bbw\c[c]} = \sum_{n=1}^{N} \frac{e^{ {\hat\bbs_n}^{(c)\top} \hat\bbw\c[c]}  {\hat\bbs_n^{(c)}} {\hat\bbs_n}^{(c)\top}}{(1+e^{ {\hat\bbs_n}^{(c)\top} \hat\bbw\c[c]})^2}.
\end{align}
Since $\frac{e^{ {x} }}{(1+e^{x})^2} \leq \frac{1}{4}$, the majorization matrix for $\hat \bbw\c[c]$ can be chosen as $\bbM_{\hat \bbw\c[c]}= \frac{1}{4} \text{diag}\{\sum_{n=1}^{N} |{\hat \bbs_n^{(c)}}| |{\hat \bbs_n^{(c)\top}}| {\bf 1}_{(K+1) \times 1} \} $ with step size $\delta \bbI$.  $\nu_{\hat \bbw\c[c]}$ is then defined as
\begin{align}
	&\nu_{\hat\bbw\c[c]} = \tilde\bbw\c[c] + \nonumber \\
    &\bbM_{\hat\bbw\c[c]}^{-1}\sum_{n=1}^N \left[(1-y_n\c[c]) \hat\bbs_n\c[c] - \frac{e^{{\hat\bbs_n}^{(c)\top} \tilde\bbw\c[c]}}{1 + e^{{\hat\bbs_n}^{(c)\top} \tilde\bbw\c[c]}} \hat\bbs_n\c[c] \right]
\end{align}
where $\tilde\bbw\c[c]$ is obtained from line (5) in Algorithm \ref{alg:bpgm}, based on the previous $\hat\bbw\c[c]$. The last step of BPG-M directly updates $\hat \bbw\c[c]$ as $\nu_{\hat \bbw\c[c]}$.The overall algorithm to solve problem (\ref{costfun}) is listed in table~\ref{table:alg2}.

\begin{table*}[t]
	\centering
        \caption{\normalfont MIML-CDL Training Algorithm Details. Table \ref{table:alg2} outlines the iterative training procedure for our MIML-CDL model, which jointly updates dictionaries, sparse codes, and classifier weights using a block coordinate optimization framework. Each subproblem is efficiently solved using majorization techniques and proximal algorithms (e.g., BPGM, ADMM), ensuring convergence while handling both reconstruction and label supervision. This unified approach enables the model to capture shared and class-specific structures in multi-instance multi-label settings.} 
	\begin{tabular}{l}
		\hline
		Initialize $\mathcal{\bar D}, \bar \bbS_n, \bbW, \bbb$ with random values \\
		\hline
		\textbf{while} not converge: \\
		\hspace{0.5cm}/* Update $\mathcal{D}\c[0]$ */ \\
		\hspace{0.5cm}\textbf{for} all $\dvec_k\c[0]$ in $\mathcal{D}\c[0]$: \\
		\hspace{1cm} Calculate truncated Toeplitz matrix $\bbT_{s_{n,k}\c[0]}$ by (\ref{Tsnk0}) \\
		\hspace{1cm} Calculate majorization matrix $\bbM_{\dvec_k\c[0]} = diag\{\sum_{n=1}^{N} 4|\bbT_{s_{n,k}\c[0]}|^\top |\bbT_{s_{n,k}\c[0]}|{\bf 1}_{M \times 1} \}$, step size is $\delta \bbI$ \\
		\hspace{1cm} Using Algorithm \ref{alg:bpgm} to solve $\min_{\dvec_k\c[0]} \sum_{n=1}^{N} \frac{1}{2} \|2\bbx_n - \alpha_n - \beta_n - 2\bbT_{s_{n,k}\c[0]} \dvec_k\c[0] \|_2^2 + \mu \|\mathcal{D}\c[0]\|_*, \ \ s.t. \ \|\dvec_k\c[0]\|\leq 1$ \\
		\hspace{1cm} The last step of Algorithm \ref{alg:bpgm} is to solve $\min_{\dvec_k\c[0]} \frac{1}{2} \| \dvec_k\c[0] - \nu_{\dvec_k\c[0]} \|^2_{\bbM_{\dvec_k\c[0]}} + N\mu \|\mathcal{D}\c[0]\|_*, \ \ s.t. \ \|\dvec_k\c[0]\|\leq 1.$ \\
		\hspace{1cm} /* solve the last step by ADMM equivalent to solve (\ref{admm}) */ \\
		\hspace{1cm}\textbf{while} not converge: \\
		\hspace{1.5cm} 1. update $\dvec_k\c[0]$ through $\min_{\dvec_k\c[0]} \frac{1}{2} \| \dvec_k\c[0] - \nu_{\dvec_k\c[0]} \|^2_{\bbM_{\dvec_k\c[0]}} + \frac{\rho}{2} \|\bbz_k-\dvec_k\c[0]+ \frac{1}{\rho}\bby_k\|_2^2$ using QCQP \\
		\hspace{1.5cm} 2. update $\bbO$ through $N\mu \|\bbO\|_* + \frac{\rho}{2}\|\bbO - \mathcal{D}\c[0]+\frac{1}{\rho}\bbY\|_F^2$ using SVT \\
		\hspace{1.5cm} 3. update $\bbY$, simply by $\bbY = \bbY + \rho (\bbO-\mathcal{D}\c[0])$, where $\rho > 1$. \\
		\hspace{1.5cm} 4. Check convergence $\|\bbO - \mathcal{D}\c[0]\| \leq \epsilon$ \\
		\hline
		\hspace{0.5cm}/* Update $\mathcal{D}$ */ \\
		\hspace{0.5cm}\textbf{for} all $\dvec_k\c[c]$ in $\mathcal{D}$: \\
		\hspace{1cm} Calculate truncated Toeplitz matrix $\bbT_{s_{n,k}\c[c]}$ by (\ref{Tsnkc}) \\
		\hspace{1cm} Calculate majorization matrix $\bbM_{\dvec_k\c[c]} = diag\{\sum_{n=1}^{N} |\bbT_{s_{n,k}\c[c]}|^\top |\bbT_{s_{n,k}\c[c]}|{\bf 1}_{M \times 1} \}$, step size is $\delta \bbI$ \\
		\hspace{1cm} Using Algorithm \ref{alg:bpgm} to solve $\min_{\dvec_k\c[c]} \sum_{n=1}^{N} \frac{1}{2} \| \bbT_{s_{n,k}\c[c]} \dvec_k\c[c] -\gamma_n \|_2^2, \ \ s.t. \ \|\dvec_k\c[c]\|\leq 1$ \\
		\hspace{1cm} The last step of Algorithm \ref{alg:bpgm} is to solve $\min_{\dvec_k\c[c]} \frac{1}{2} \| \dvec_k\c[c] - \nu_{\dvec_k\c[c]} \|^2_{\bbM_{\dvec_k\c[c]}}, \ \ s.t. \ \|\dvec_k\c[c]\|\leq 1$ by accelerated Newton's method. \\
		\hline
		\hspace{0.5cm}/* Update $\bbS_n\c[0]$ */ \\
		\hspace{0.5cm}\textbf{for} all $\bbs_{n,k}\c[0]$ in $\bbS_n\c[0]$: \\
		\hspace{1cm} Calculate truncated Toeplitz matrix $\bbT_{d_k\c[0]}$ by (\ref{Td0}) \\
		\hspace{1cm} Calculate majorization matrix $\bbM_{\bbS\c[0]} = diag\{ 4|\bbT_{d_k\c[0]}|^\top |\bbT_{d_k\c[0]}|{\bf 1}_{T \times 1}\}$, step size is $\delta \bbI$ \\
		\hspace{1cm} Using Algorithm \ref{alg:bpgm} to solve $\min_{\bbs_{n,k}\c[0]} \frac{1}{2} \|2\bbx_n - \alpha_n - \beta_n - 2\bbT_{d_k\c[0]} \bbs_{n,k}\c[0] \|_2^2+ \lambda \|\bbs_{n,k}\c[0] \|_1$ \\
		\hspace{1cm} The last step of Algorithm \ref{alg:bpgm} is to solve $\min_{\bbs_{n,k}\c[0]} \frac{1}{2} \|\bbs_{n,k}\c[0] - \nu_{\bbs_{n,k}\c[0]} \|_{\bbM_{\bbS\c[0]}}^2+ \lambda \|\bbs_{n,k}\c[0] \|_1,$ by thresholding operator. \\
		\hline
		\hspace{0.5cm}/* Update $\bbS_n$ */ \\
		\hspace{0.5cm}\textbf{for} all $\bbs_{n,k}\c[c]$ in $\bbS_n\c[c]$: \\
		\hspace{1cm} Calculate truncated Toeplitz matrix $\bbT_{d_k\c[c]}$ by (\ref{Tdck}) \\
		\hspace{1cm} Calculate majorization matrix $\bbM_{\bbS\c[c]} = diag\{|\bbM_{\bbS}|^\top |\bbM_{\bbS}|{\bf 1}_{T \times 1}\}$, where $\bbM_{\bbS}= [ 2\bbT_{d_k\c[c]}^\top, \frac{1}{2}\sqrt{\eta} \bbp |w_k\c[c]|]^\top $, step size is $\delta \bbI$ \\
		\hspace{1cm} Using Algorithm \ref{alg:bpgm} to solve $\min_{\bbs_{n,k}\c[c]} 2 \| \bbT_{d_k\c[c]} \bbs_{n,k}\c[c] -\gamma_n \|_2^2 + \lambda \|\bbs_{n,k}\c[c] \|_1 + \eta (1-y_n\c[c])(\bbp^\top \bbs_{n,k}\c[c] w_k\c[c]) + \eta \log (1+ e^{({\hat\bbs_n}^{(c)\top} \hat \bbw\c[c])}).$ \\
		\hspace{1cm} The last step of Algorithm \ref{alg:bpgm} is to solve $\min_{\bbs_{n,k}\c[c]} \frac{1}{2} \|\bbs_{n,k}\c[c] - \nu_{\bbs_{n,k}\c[c]} \|_{\bbM_{\bbS\c[c]}}^2+ \lambda \|\bbs_{n,k}\c[c] \|_1,$ by thresholding operator. \\
		\hline
		\hspace{0.5cm}/* Update $\bbW$ and $\bbb$ */ \\
		\hspace{0.5cm}\textbf{for} all $\bbw\c[c]$ in $\bbW$: \\
		\hspace{1cm} Calculate majorization matrix $\bbM_{\hat \bbw\c[c]}= \frac{1}{4} diag\{\sum_{n=1}^{N} |{\hat \bbs_n}\c[c]| |{\hat \bbs_n}^{(c)\top}| {\bf 1}_{(K+1) \times 1} \} $, step size is $\delta \bbI$ \\
		\hspace{1cm} Using Algorithm \ref{alg:bpgm} to solve $\min_{\hat \bbw\c[c]} \sum_{n=1}^{N} -(1-y_n\c[c])({\hat \bbs_n}^{(c)\top} \hat \bbw\c[c]) + \log(1+ e^{ {\hat \bbs_n}^{(c)\top} \hat \bbw\c[c]}).$ \\
		\hline
		\hspace{0.5cm}/* Check convergence */ \\
		\hspace{0.5cm}\textbf{if} the overall loss changing rate $\leq \epsilon$, break \\
		\hline
	\end{tabular}
	\label{table:alg2}
\end{table*}
\section{Extension to Tensor Data}
\label{sec:ext}

In previous sections, we represented the $n$-th bag of data as $\bbx_n\in \mathbb{R}^T$, where each instance is a $1$-D vector and the bag of instances is concatenated sequentially. For the proposed MIML-CDL model (\ref{costfun}), the $k$-th atom in the $c$-th class is $\bbd_k\c[c] \in \mathbb{R}^M$ with its corresponding sparse coefficient $\bbs_{n,k}\c[c] \in \mathbb{R}^T$ for the $n$-th bag in the $c$-th class.
Although high-dimensional data can be vectorized into $1$-D representations, such flattening causes significant loss of structural information during convolution operations. For instance, a $2$-D image processed with a $2$-D convolution kernel naturally captures both vertical and horizontal patterns. In contrast, when the same image is vectorized horizontally, a $1$-D convolution can only extract horizontal patterns, completely missing vertical relationships. This structural information loss due to vectorization leads to degraded classification performance.
Therefore, extending our proposed model to handle high-dimensional data in its native format is essential for preserving spatial relationships and achieving optimal classification results.

\subsection{Tensor data}
The proposed model naturally extends to higher-dimensional data through tensor formulations. Consider first a 2D bag representation $\bbX_n\in \mathbb{R}^{F \times T}$, where instances are temporally concatenated along the second dimension (alternative instance configurations will be discussed in Section~\ref{sec:sep_ins}). Let the $j$-th instance in the $n$-th bag be $\bbI_{n,j} \in \mathbb{R}^{F \times T_j}$ with $\sum_j T_j = T$. Each dictionary atom for class $c$ becomes a 2D filter $\mathcal{D}_k\c[c] \in \mathbb{R}^{F \times M}$, while the corresponding sparse coefficients $\bbs_{n,k}\c[c] \in \mathbb{R}^T$ retain their original dimensionality.

This formulation generalizes to arbitrary dimensions through tensor notation. For an $N$-dimensional tensor bag $\overrightarrow{\bbX_n} \in \mathbb{R}^{T_1 \times \cdots \times T_N}$, the $c$-th class dictionary atoms become $N$-dimensional tensors $\overrightarrow{\mathcal{D}_k\c[c]} \in \mathbb{R}^{T_1 \times \cdots \times T_{N-1} \times M}$, while the sparse coefficients $\bbs_{n,k}\c[c] \in \mathbb{R}^T$ maintain their original vector form. Crucially, the coefficient dimensionality remains invariant to the input dimension, preserving the pooling operation's compatibility.

For implementations, we note that current majorization matrix computations support up to $2$-D operations. Higher-dimensional data ($3$-D+) can be handled by preserving two principal dimensions and collapsing remaining dimensions through reindexing. Formally, this amounts to reshaping a tensor $\overrightarrow{\bbX} \in \mathbb{R}^{T_1 \times \cdots \times T_D}$ into a matrix $\bbX' \in \mathbb{R}^{T_1 \times (\prod_{d=2}^D T_d)}$, effectively decomposing the high-dimensional structure into separate $2$-D slices while preserving the model's core functionality.

\subsection{Separated instances in a bag}
\label{sec:sep_ins}
When handling separated instances in a bag, two distinct scenarios arise: 1) Instances may be small patches extracted from a larger bag of data (e.g., a single image containing both cats and dogs, where individual animal regions constitute separate instances), or 2) Instances may be independent data elements (e.g., a collection of separate images, each representing one instance in the bag).

For the first scenario, our model must adaptively adjust the dimensions of atoms and sparse coefficients. Given a 2-dimensional bag $\bbX_n\in \mathbb{R}^{F \times T}$, the $j$-th instance in the $n$-th bag becomes $\bbI_{n,j} \in \mathbb{R}^{F_j \times T_j}$, where $\sum_j F_j \leq F$ and $\sum_j T_j \leq T$. Consequently, each atom in the $c$-th class is defined as ${\mathcal{D}_k\c[c]} \in \mathbb{R}^{\max(\{F_j\}_j) \times \max(\{T_j\}_j)}$, with corresponding sparse coefficients $\bbs_{n,k}\c[c] \in \mathbb{R}^{FT}$. The model structure remains unchanged except for adjustments to the dimensionality of the pooling operator and majorization matrix.

In the second scenario, where instances are independent data elements, the $n$-th example is represented as $\bbX_n = \{\bbI_{n,j}\}_{j=1}^J$, with each instance $\bbI_{n,j} \in \mathbb{R}^{F \times T}$. The atoms for the $c$-th class maintain their form as $\mathcal{D}_k\c[c] \in \mathbb{R}^{F \times M}$, while the sparse coefficients are denoted as $\bbs_{n,k,j}\c[c] \in \mathbb{R}^T$ for the $n$-th bag, $c$-th class, $k$-th atom, and $j$-th instance. The computational pipeline remains consistent with our original formulation, with the key modification being that the pooling operation now operates on the concatenation of all $J$ instance coefficients $\bbs_{n,k,j}\c[c]$.

\section{Numerical Experiments}
\label{sec:results}
In this section, we evaluate our proposed algorithm on both synthetic and real-world datasets to demonstrate its effectiveness. We implemented our approach with support for both CPU and GPU acceleration, and the complete source code is available on GitHub\footnote{https://github.com/chenhao1umbc/WSCDL}. We present comprehensive results, including classification performance metrics, detailed numerical analysis, and comparisons against state-of-the-art related works.

\subsection{Synthetic data}
We constructed a synthetic one-dimensional dataset comprising four discriminative classes in addition to a background category. Each class and the background are characterized by five prototypical features (30 samples per feature), generated using sinusoidal and sawtooth waveforms. These features manifest as burst events—up to five consecutive repetitions—interspersed with variable-length zero-padded intervals. Each data instance is a time series of 1600 samples, containing randomly positioned bursts.

The dataset includes the full set of 15 possible multi-label combinations derived from the four classes. The test set comprises 50 examples for each combination (750 examples in total), while the training set (550 examples) intentionally excludes single-label instances to encourage the algorithm to learn in the absence of single-label supervision. To simulate real-world interference and assess robustness, we added white Gaussian noise at a signal-to-noise ratio (SNR) of 10 dB. Class labels are encoded as binary vectors indicating the presence or absence of each class.

\begin{figure}[t]
	\centering
	\hspace*{-3pt}\includegraphics[width=0.99\columnwidth]{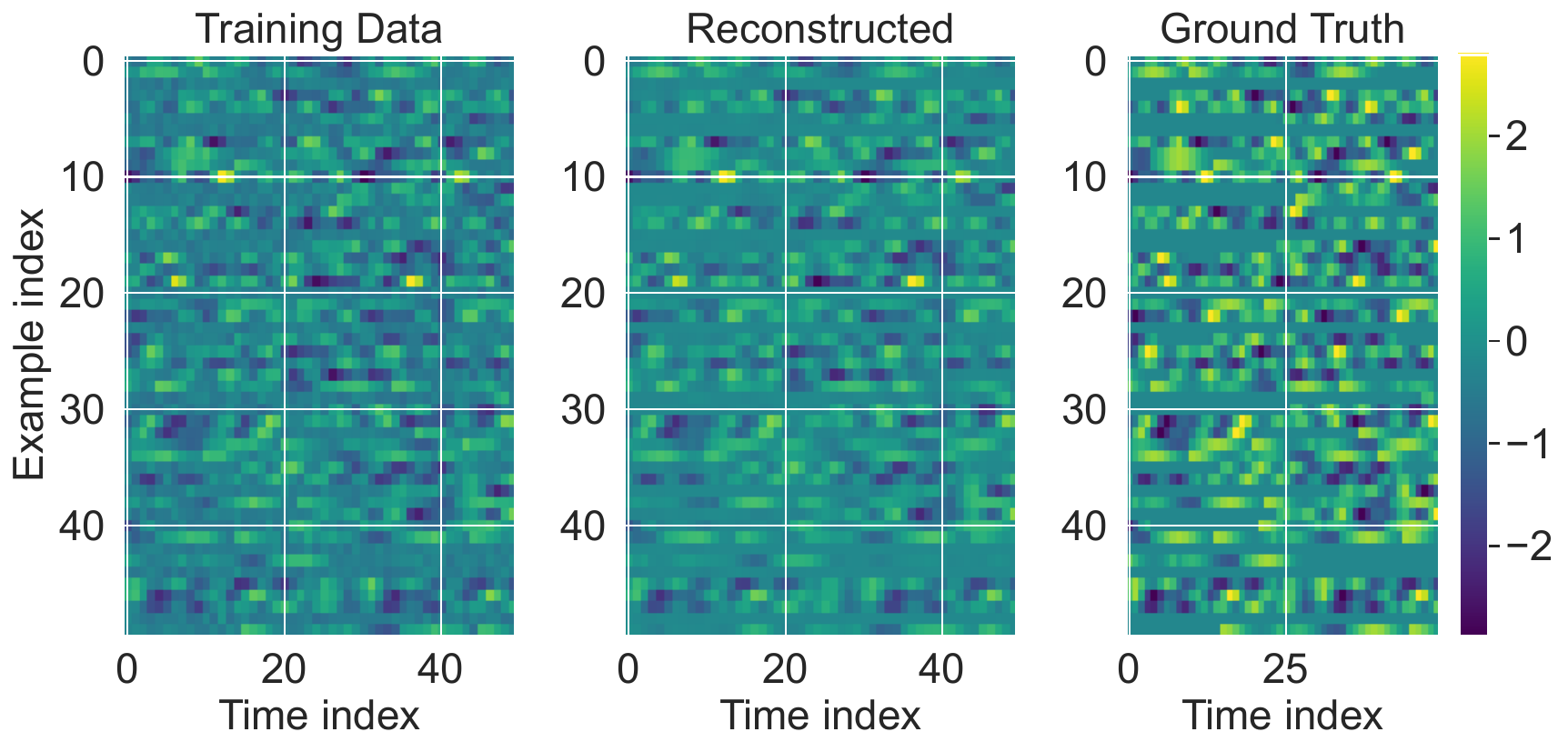}\hspace*{-3pt}
	\caption{Visualization of the first 50 samples from the first 50 training examples. Each panel shows a heatmap with time indices on the x-axis and example indices on the y-axis. (Left) Noisy training data. (Middle) Data reconstructed by the model, with noise effectively removed. (Right) Ground-truth data without noise. The color scale represents signal amplitude.}
	\label{fig:data}
	\vspace{-5mm}
\end{figure}

Fig.~\ref{fig:data} visualizes the first 50 time samples of the first 50 training examples as heatmaps, with time indices along the x-axis and example indices along the y-axis. The left panel shows the noisy training data, the middle panel displays the reconstructed data produced by our model, and the right panel presents the corresponding ground-truth data without noise. The close resemblance between the reconstructed and ground-truth panels demonstrates the model's ability to effectively denoise and recover the underlying structure of the data.

\begin{figure}[t]
	\centering
	\subfigure[Feature 1]{\label{fig:feat1}\includegraphics[width=0.49\columnwidth ]{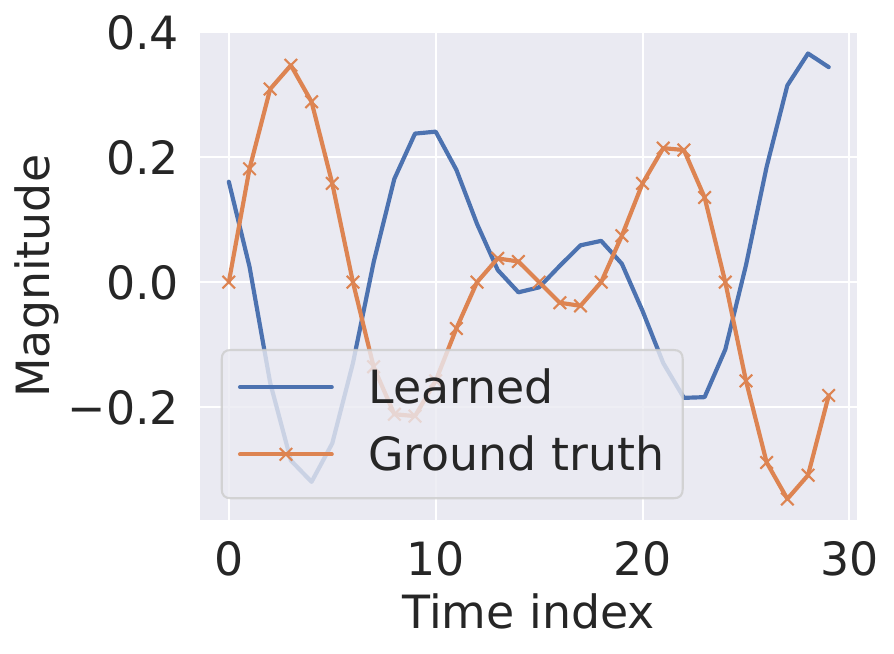}}
	\subfigure[Feature 2]{\label{fig:feat2}\includegraphics[width=0.49\columnwidth ]{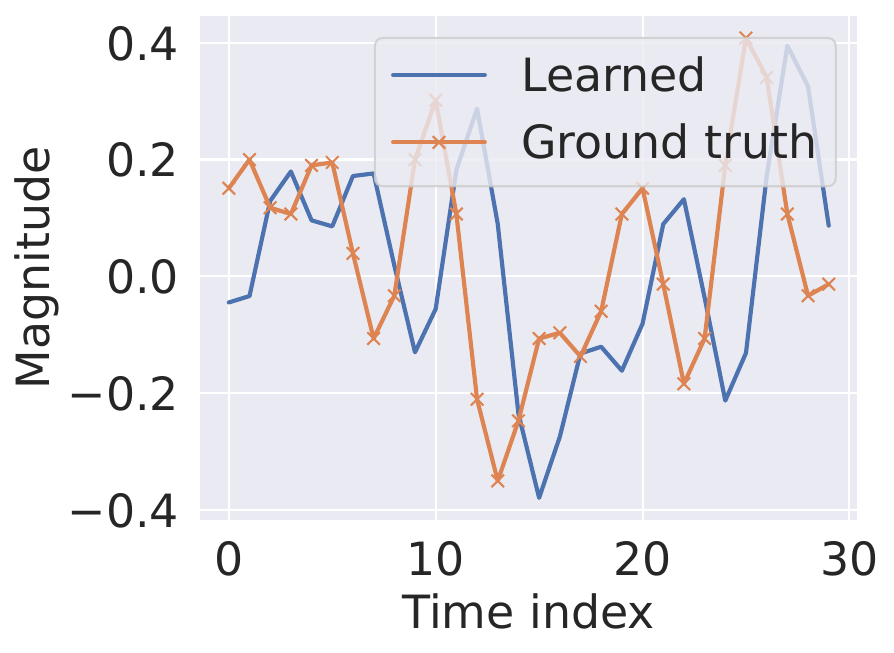}}
	\subfigure[Feature 3]{\label{fig:feat3}\includegraphics[width=0.49\columnwidth ]{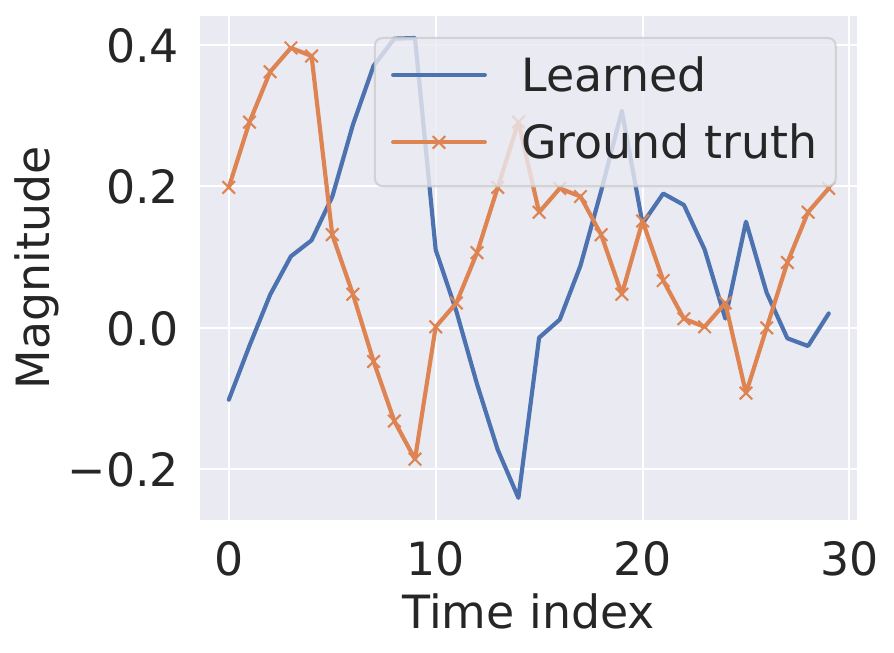}}
	\subfigure[Feature 4]{\label{fig:feat4}\includegraphics[width=0.49\columnwidth ]{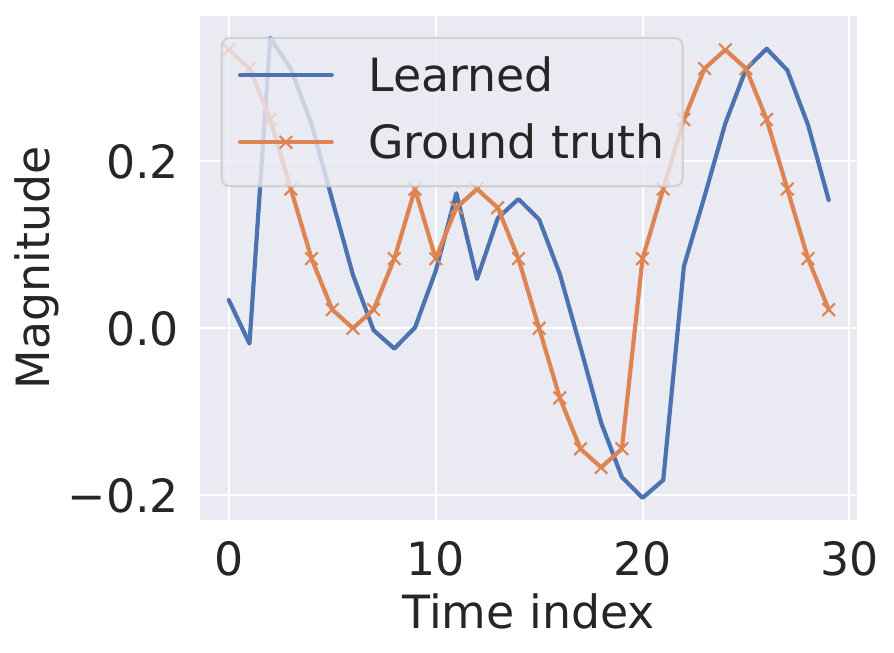}}
	\subfigure[Common feature]{\label{fig:comm_feat}\includegraphics[width=0.49\columnwidth ]{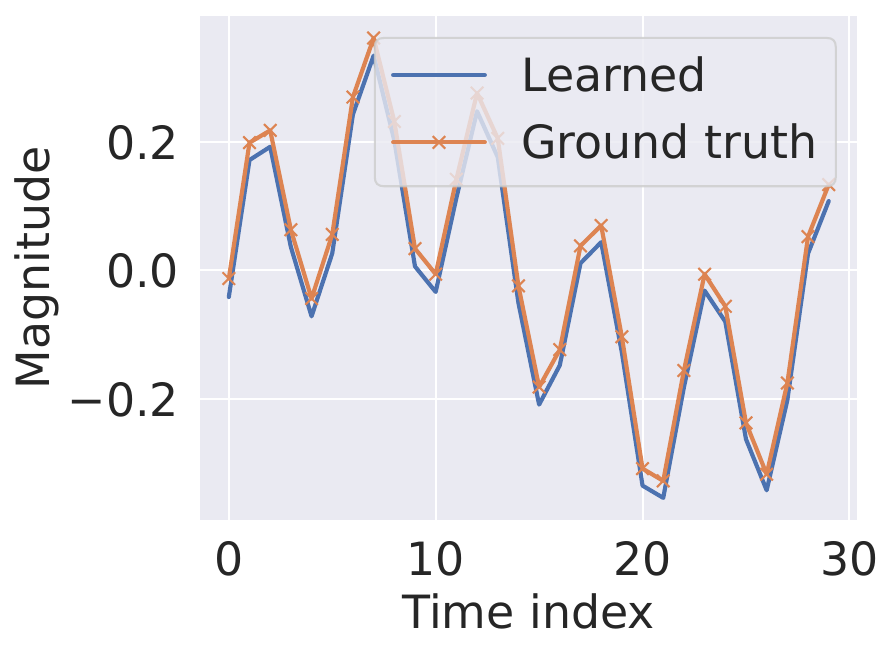}}
	\subfigure[Learning curve]{\label{fig:loss_func}\includegraphics[width=0.49\columnwidth ]{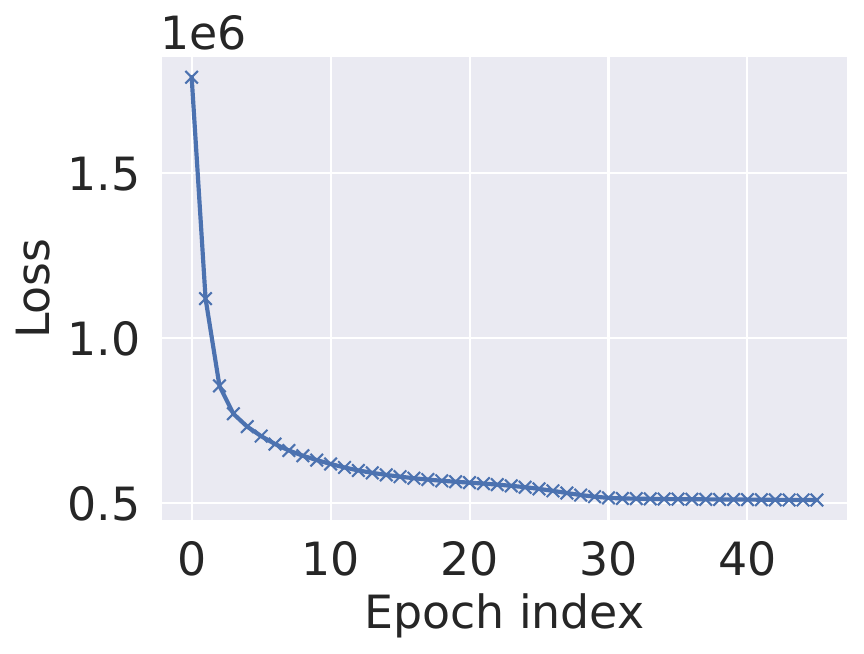}}

	\caption{
		Dictionary atoms recovery and optimization convergence results.
		(a)-(e): Each plot compares a learned dictionary atom (blue) with its corresponding ground-truth feature (orange, with "x" markers) for the four classes and the common (background) feature. The learned features closely match the ground truth, with minor time shifts due to burst structure and data truncation. These shifts demonstrate the translation invariance property of our convolutional approach, which allows the model to recognize patterns regardless of their exact position.
		(f): Training loss versus epoch index, showing rapid and stable convergence of the proposed algorithm.
	}
	\label{fig:feat}
\end{figure}

To visualize of learned features and training performance, Fig.~\ref{fig:feat1}-~\ref{fig:comm_feat} compare our model's learned dictionary atoms (blue lines) against the ground-truth features (orange lines with "x" markers) for each of the four classes and the background. Although the learned features accurately capture the essential patterns of the ground-truth signals, we observe temporal displacements in some cases, for example, feature 1 exhibits a 15-sample shift and feature 4 shows a 2-sample shift. These circular shifts arise naturally from two aspects of our data generation process: the repetitive burst structure of the features and the finite window truncation in the synthetic data. Importantly, these shifts do not compromise reconstruction quality because convolution operations inherently possess translation invariance, allowing the model to effectively recognize patterns regardless of their exact location. Fig.~\ref{fig:loss_func} demonstrates that the loss decreases rapidly during the first 10 epochs and continues to decrease more gradually, stabilizing at a local minimum around epoch 40. This behavior indicates that the proposed optimization algorithm achieves both efficient and stable convergence.

\begin{figure}[t]
	\centering
	\subfigure[True labels]{\label{fig:true_label}\includegraphics[width=0.49\columnwidth ]{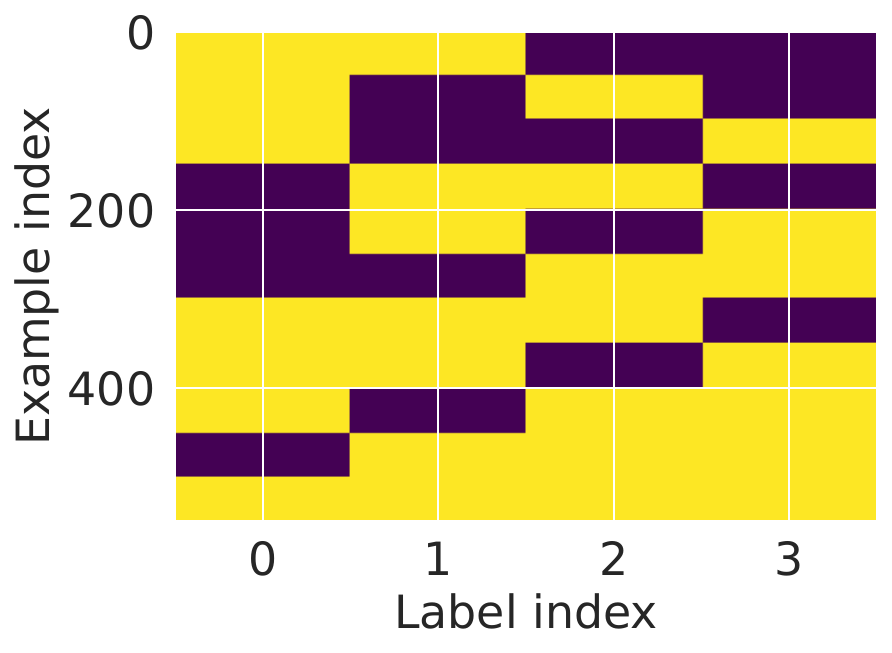}}
	\subfigure[Reconstructed labels]{\label{fig:rec_label}\includegraphics[width=0.49\columnwidth ]{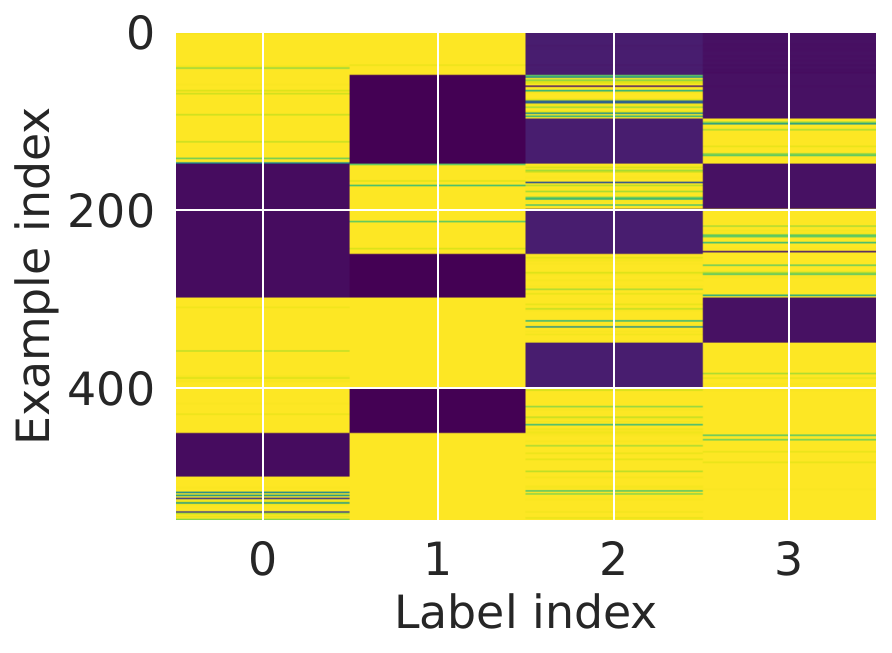}}

	\subfigure[Reconstr. labels with threshold]{\label{fig:rec_label_thr}\includegraphics[width=0.49\columnwidth ]{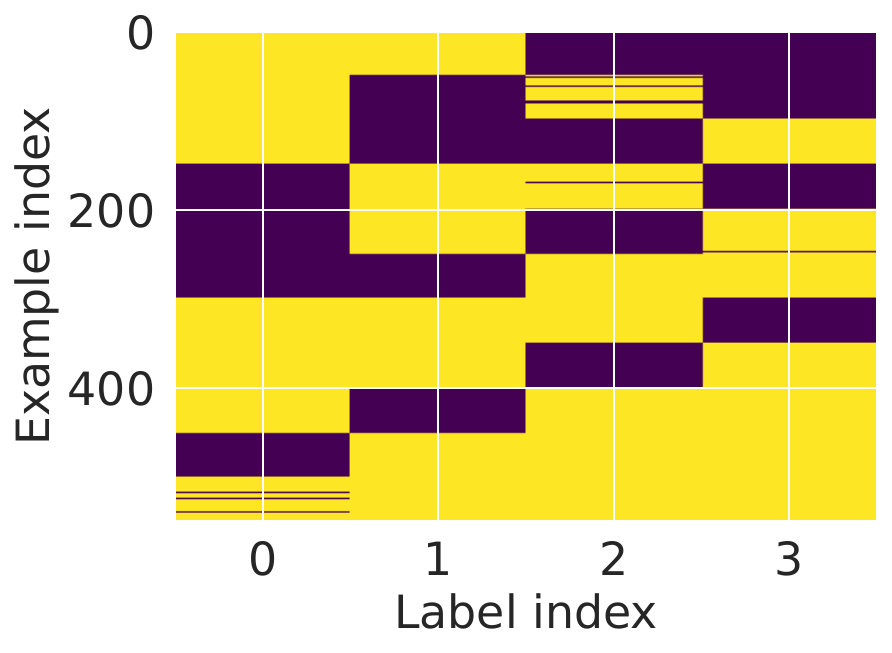}}
	\subfigure[Sparse coeff. with threshold]{\label{fig:sparse_coeffs_thr}\includegraphics[width=0.49\columnwidth ]{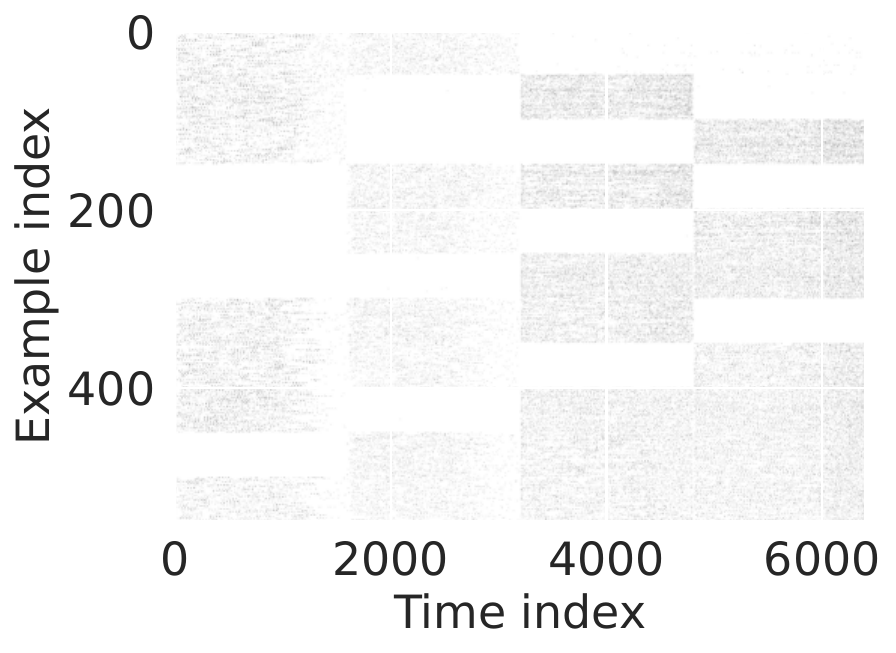}} 

	\caption{(a) The left figure is the ground-truth of the labels. (b) The middle figure is the reconstructed labels. (c) The right figure is the reconstructed label with threshold $=0.5$. (d) The figure shows the sparse coefficients with threshold.}
	\label{fig:labels}
\end{figure}

Fig.~\ref{fig:labels} presents the classification results on the test data set. Each subfigure plots class indices (x-axis) versus test example indices (y-axis), with color intensity indicating class presence. Fig.~\ref{fig:true_label} shows ground truth labels, while Fig.~\ref{fig:rec_label} displays the model's raw predictions. After applying a threshold of 0.5 to obtain binary classification decisions are shown in Fig.~\ref{fig:rec_label_thr}. Fig.~\ref{fig:sparse_coeffs_thr} visualizes the thresholded sparse coefficients that form the basis of these predictions. Notably, although the model was not exposed to single-label examples during training, it accurately identifies such cases during testing. The resulting classification accuracy reaches $0.9827$, underscoring the strong generalization capability of the model in weakly supervised settings.

\begin{figure}[t]
	\vspace{-20mm}
	\centering
	\includegraphics[width=0.99\columnwidth ]{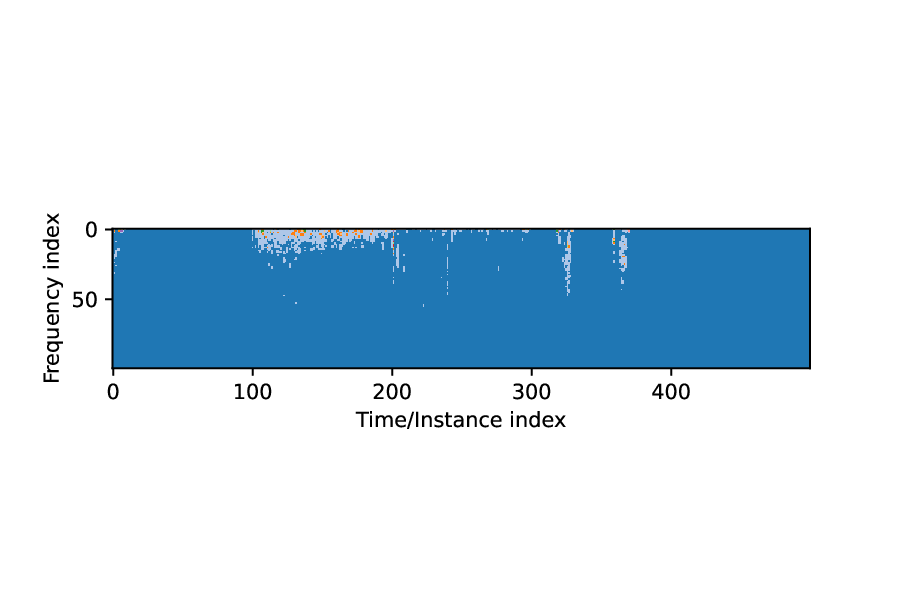}
	\vspace{-20mm}
	\caption{One training example of ESC-10 data of $5$ concatenated instances, respectively with labels as ``Sneezing", ``Sea waves", ``Cracking fire", ``Sneezing" and ``Null".}
	\label{fig:esc}
	\vspace{-5mm}
\end{figure}

\subsection{Public real-world data}
We evaluate our method on ESC-10, a curated subset of the ESC-50 dataset \cite{K2015} containing 10 environmental sound classes (Chainsaw, Clock tick, Cracking fire, Crying baby, Dog, Helicopter, Rain, Rooster, Sea waves, and Sneezing). Each recording is a 5-second audio clip sampled at 44.1kHz. To create a weakly supervised setting, we organized 5,000 instances into 1,000 bags with 5 instances per bag, where each instance could be either a sound from ESC-10 or null. We transformed the audio data into $100 \times 100$ time-frequency representations using Short-Time Fourier Transform (STFT) and downsampling. Fig.~\ref{fig:esc} shows an example bag with five concatenated instances (indices 0-99, 100-199, 200-299, 300-399, and 400-499) labeled as ``Sneezing," ``Sea waves," ``Cracking fire," ``Sneezing," and ``Null" respectively. The y-axis represents frequency (with lower frequencies at the top), and we display only the one-sided spectrum due to the symmetry of the STFT.

\begin{table*}[htbp]
	\centering
	\caption{\normalfont Comparison of classification performance metrics across different weakly supervised methods on the ESC-10 dataset. Best results are highlighted in \textbf{bold}.}
	\begin{tabular}{lccccc}
		\hline
		                  & Accuracy        & Recall          & Precision       & F1 score        & \textbf{AUC}    \\
		\hline
		Deep MIML         & 0.5432          & 0.2918          & 0.4029          & 0.3321          & 0.5119          \\
		MIML-kNN          & 0.5307          & \textbf{0.6035} & 0.3927          & 0.4719          & 0.5230          \\
		MIML-SVM (linear) & 0.6247          & 0.5579          & \textbf{0.5062} & 0.5307          & 0.6639          \\
		MIML-SVM (RBF)    & 0.6207          & 0.4950          & 0.5037          & 0.4991          & 0.6628          \\
		WSDL              & 0.5405          & 0.2956          & 0.3837          & 0.3878          & 0.5050          \\
		\rowcolor[gray]{0.8}
		WSCDL(ours)       & \textbf{0.6330} & 0.5506          & 0.5321          & \textbf{0.5423} & \textbf{0.6753} \\
		\hline
	\end{tabular}
	\vspace{-1mm}
	\label{tab:auc}
\end{table*}

The comparison of performance is conducted among several weakly supervised methods: WSDL~\cite{YRF18}, our proposed WSCDL, MIML-SVM~\cite{ZZ2007,CL2011}, MIML-kNN~\cite{MZ2010}, and Deep MIML network~\cite{FZ17}. We divided the dataset into 800 bags as training/validation examples and 200 bags for testing. For validation, we employed 5-fold cross-validation with 160 bags per fold. For MIML-SVM and MIML-kNN, which require separated instances in each bag, we followed the scenario 2 described in Section~\ref{sec:sep_ins}, while Deep MIML network corresponds to scenario 1.

For this multi-label classification task, we evaluated performance using accuracy, recall, precision, F1-score, and Area Under Curve (AUC). After hyperparameter tuning, MIML-kNN performed best with $k=7$. For MIML-SVM, we tested both linear and RBF kernel variants. The Deep MIML implementation used a reduced VGG architecture~\cite{SZ2015} (1/5 of the original depth) with batch size 64. WSDL was configured with window size 100, sparsity constraint $N=100$, $\lambda=0.1$, and 2 atoms per class. Our WSCDL method used window size 95, $\lambda=0.1$, $\eta=0.01$, $\mu=0.1$, with average pooling and cross-entropy loss. Table~\ref{tab:auc} presents the averaged results from 5 independent test runs.
The performance differences stem from each method's alignment with the problem structure. For fair comparison, we used a dynamic threshold (mean of maximum and minimum predicted probabilities) instead of a fixed 0.5 threshold. WSDL, Deep MIML, and MIML-kNN performed near random guessing due to several limitations: MIML-kNN struggles with high intra-class variance; Deep MIML suffers from insufficient training data and mismatched model assumptions; WSDL lacks our crucial $\bbW$ term for projecting instance labels to bag labels. Our method's incorporation of $\mathcal{D}_0$ effectively filters background noise, enhancing performance.

MIML-SVM and our WSCDL approach significantly outperformed other methods. MIML-SVM aligned well with the problem structure, though interestingly, its nonlinear kernel variant showed no improvement over the linear version. Our algorithm achieved the highest accuracy ($0.6330$) among all the compared methods, despite the inherent challenges of weakly supervised learning on this complex dataset.

\begin{figure}[h]
	\centering
	\includegraphics[width=0.99\columnwidth ]{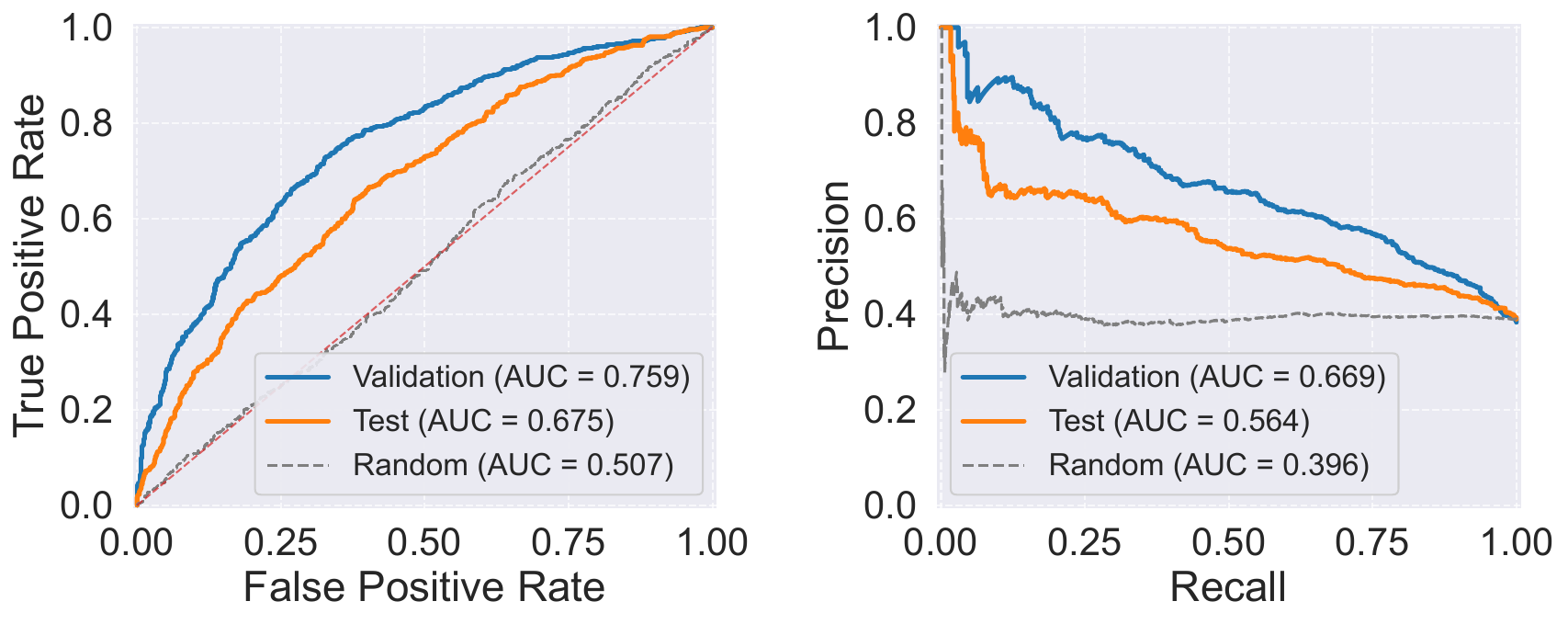}\vspace{-5mm}
	\caption{Performance evaluation on the ESC-10 dataset. (Left) Receiver Operating Characteristic curves with the blue curve showing the median result from 5-fold validation (AUC=0.759) and the orange curve representing the median test result (AUC=0.675). The diagonal line indicates random guessing (AUC=0.507). (Right) Precision-Recall curves with validation (AUC=0.669) and test (AUC=0.564) results, compared to random performance (AUC=0.396).}
	\label{fig:roc}
\end{figure}

Fig.~\ref{fig:roc} quantifies the performance of our classifier in the ESC-10 data set through two complementary visualization approaches. The left panel presents Receiver Operating Characteristic (ROC) curves that plot the True Positive Rate against the False Positive Rate. For statistical robustness, we display the median-performing instance from our 5-fold validation experiments (blue curve, AUC=0.759) alongside the median result from five independent test evaluations (orange curve, AUC=0.675). The diagonal green reference line represents the random classification performance (AUC=0.507). The right panel shows the corresponding Precision-Recall (PR) curves that highlight our model's trade-off between precision and recall, with validation (AUC=0.669) and test (AUC=0.564) compared against random classification (AUC=0.396). The consistent performance gap between validation and test metrics reflects an expected challenge in weakly supervised paradigms, specifically, the propagation of uncertainty from bag-level to instance-level labels. Notably, this performance differential was evident across all benchmark methods in our comparative analysis, confirming that our approach achieves competitive discriminative capability despite the intrinsic difficulties of the weakly labeled ESC-10 acoustic classification task.
\section{Conclusion} \label{sec:conclusion}
This paper introduced a novel CDL framework for weakly supervised MIMLclassification. Our approach demonstrates superior performance when training data is limited or when individual instances lack clear labels. Key innovations include: (1) a shared background dictionary that enhances the discriminative capacity of class-specific features, (2) an efficient Block Proximal Gradient method with Majorization that ensures fast convergence despite the computational demands of convolution operations, and (3) a projection mechanism that significantly improves bag-level label prediction accuracy. While our model conceptually extends to tensor data, current effective implementation is limited to $2$-D optimization due to the challenge of selecting majorization matrix constraints in tensor data. Future work will focus on improving instance-level label predictions and extending the optimization method to handle higher-dimensional tensor data directly.

\bibliographystyle{IEEEtran}
\bibliography{ref.bib}

\begin{thebibliography}{10}
\providecommand{\url}[1]{#1}
\csname url@samestyle\endcsname
\providecommand{\newblock}{\relax}
\providecommand{\bibinfo}[2]{#2}
\providecommand{\BIBentrySTDinterwordspacing}{\spaceskip=0pt\relax}
\providecommand{\BIBentryALTinterwordstretchfactor}{4}
\providecommand{\BIBentryALTinterwordspacing}{\spaceskip=\fontdimen2\font plus
\BIBentryALTinterwordstretchfactor\fontdimen3\font minus \fontdimen4\font\relax}
\providecommand{\BIBforeignlanguage}[2]{{%
\expandafter\ifx\csname l@#1\endcsname\relax
\typeout{** WARNING: IEEEtran.bst: No hyphenation pattern has been}%
\typeout{** loaded for the language `#1'. Using the pattern for}%
\typeout{** the default language instead.}%
\else
\language=\csname l@#1\endcsname
\fi
#2}}
\providecommand{\BIBdecl}{\relax}
\BIBdecl

\bibitem{AEB06}
M.~Aharon, M.~Elad, and A.~Bruckstein, ``K-svd: An algorithm for designing overcomplete dictionaries for sparse representation,'' \emph{IEEE Trans. Signal Process.}, vol.~54, no.~11, pp. 4311--4322, 2006.

\bibitem{BDE09}
A.~Bruckstein, D.~Donoho, and M.~Elad, ``From sparse solutions of systems of equations to sparse modeling of signals and images,'' \emph{SIAM Review}, vol.~51, no.~1, pp. 34--81, 2009.

\bibitem{XY16}
Y.~Xu and W.~Yin, ``A fast patch-dictionary method for whole image recovery,'' \emph{Inverse Problems and Imaging}, vol.~10, no.~2, pp. 563--583, 2016.

\bibitem{SPC14}
S.~Shekhar, V.~M. Patel, and R.~Chellappa, ``Analysis sparse coding models for image-based classification,'' in \emph{Int. Conf. on Image Process.}, 2014, pp. 5207--5211.

\bibitem{FDDL}
M.~Yang, L.~Zhang, X.~Feng, and D.~Zhang, ``Fisher discrimination dictionary learning for sparse representation,'' in \emph{Proc. Int. Conf. Computer Vis.}, Barcelona, Spain, Nov. 2011, pp. 543--550.

\bibitem{Tang19}
W.~Tang, A.~Panahi, H.~Krim, and L.~Dai, ``Analysis dictionary learning: an efficient and discriminative solution,'' in \emph{Proc. Int. Conf. Acoust. Speech Signal Process.}, May 2019, pp. 3682--3686.

\bibitem{DiCoDiLe}
I.~Dokmanic, K.~Kumanan, Y.~M. Lu, and R.~Vidal, ``Dicodile: Distributed convolutional dictionary learning,'' in \emph{IEEE Int. Conf. on Acoustics, Speech and Signal Processing (ICASSP)}, 2021, pp. 4230--4234.

\bibitem{MultiModalCDL}
R.~Liu, L.~Ma, J.~Zhang, X.~Fan, and Z.~Luo, ``Multi-modal convolutional dictionary learning,'' in \emph{Proc. of the IEEE/CVF Conf. on Computer Vision and Pattern Recognition}, 2022, pp. 11\,363--11\,372.

\bibitem{TensorCDL}
T.~G. Kolda, D.~Hong, and J.~A. Duersch, ``Tensor convolutional dictionary learning with cp low-rank activations,'' \emph{SIAM Journal on Mathematics of Data Science}, vol.~4, no.~3, pp. 1100--1125, 2022.

\bibitem{DeepM2CDL}
Y.~Zhang, Q.~Li, Y.~Guo, and D.~Zhang, ``Deepm2cdl: Deep multi-scale multi-modal convolutional dictionary learning network,'' \emph{IEEE Trans. Pattern Anal. Mach. Intell.}, vol.~45, no.~2, pp. 2287--2302, 2023.

\bibitem{chen16}
B.~Chen, J.~Li, B.~Ma, and G.~Wei, ``Convolutional sparse coding classification model for image classification,'' in \emph{2016 IEEE Int. Conference on Image Processing (ICIP)}, Sep. 2016, pp. 1918--1922.

\bibitem{jin17}
J.~Jin and C.~L.~P. Chen, ``Convolutional sparse coding for face recognition,'' in \emph{2017 4th Int. Conf. on Info., Cyber. and Comput. Social Systems (ICCSS)}, Jul. 2017, pp. 137--141.

\bibitem{YRF18}
Z.~You, R.~Raich, X.~Z. Fern, and J.~Kim, ``Weakly supervised dictionary learning,'' \emph{IEEE Trans. Signal Process.}, vol.~66, no.~10, pp. 2527--2541, May 2018.

\bibitem{ChunF18}
I.~Y. Chun and J.~A. Fessler, ``Convolutional dictionary learning: Acceleration and convergence,'' \emph{IEEE Trans. Image Process.}, vol.~27, no.~4, pp. 1697--1712, Apr. 2018.

\bibitem{CHK18}
H.~Chen, S.-J. Kim, and T.~Chatt, ``Discriminative dictionary learning for mixture component detection with application to rf signal recognition,'' in \emph{Proc. Asilomar Conf. Signals Syst. Comput.}, 2018, pp. 835--839.

\bibitem{CHK21}
H.~Chen and S.-J. Kim, ``Robust rf mixture signal recognition using discriminative dictionary learning,'' \emph{IEEE Access}, vol.~9, pp. 141\,107--141\,120, 2021.

\bibitem{GRK07}
R.~Grosse, R.~Raina, H.~Kwong, and A.~Y. Ng, ``Shift-invariant sparse coding for audio classification,'' in \emph{Proc. of the 23rd Conf. Uncertainty Artificial Intell.}, Vancouver, BC, Canada, Jul. 2007, pp. 149--158.

\bibitem{HHW15}
F.~Heide, W.~Heidrich, and G.~Wetzstein, ``Fast and flexible convolutional sparse coding,'' in \emph{Conf. on Computer Vis. and Pattern Recognition}, 2015, pp. 5135--5143.

\bibitem{ZZ2007}
Z.-H. Zhou and M.-L. Zhang, ``Multi-instance multi-label learning with application to scene classification,'' in \emph{Advances in Neural Info. Process. Syst.}, B.~Sch\"{o}lkopf, J.~Platt, and T.~Hoffman, Eds., vol.~19.\hskip 1em plus 0.5em minus 0.4em\relax MIT Press, 2007, pp. 1609--1616.

\bibitem{MZ2010}
M.~Zhang, ``A k-nearest neighbor based multi-instance multi-label learning algorithm,'' in \emph{2010 22nd IEEE Int. Conf. on Tools with Artificial Intell.}, vol.~2, 2010, pp. 207--212.

\bibitem{CL2011}
C.-C. Chang and C.-J. Lin, ``Libsvm: A library for support vector machines,'' \emph{ACM Trans. Intell. Syst. Tech.}, vol.~2, no.~3, May 2011.

\bibitem{FZ17}
J.~Feng and Z.-H. Zhou, ``Deep miml network,'' in \emph{Proc. of the Thirty-First AAAI Conf. on Artificial Intell.}, ser. AAAI'17.\hskip 1em plus 0.5em minus 0.4em\relax AAAI Press, 2017, pp. 1884--1890.

\bibitem{VuM17}
T.~H. Vu and V.~Monga, ``Fast low-rank shared dictionary learning for image classification,'' \emph{IEEE Trans. Image Process.}, vol.~26, no.~11, pp. 5160--5175, Nov. 2017.

\bibitem{BCE10}
S.~Boyd, N.~Parikh, E.~Chu, B.~Peleato, and J.~Eckstein, ``Distributed optimization and statistical learning via the alternating direction method of multipliers,'' \emph{Foundations and Trends in Machine Learning}, vol.~3, no.~1, pp. 1--122, 2010.

\bibitem{P19}
G.~Peng, ``Adaptive admm for dictionary learning in convolutional sparse representation,'' \emph{IEEE Trans. Image Process.}, vol.~28, no.~7, pp. 3408--3422, 2019.

\bibitem{KGN19}
A.~Khamparia, D.~Gupta, N.~G. Nguyen, A.~Khanna, B.~Pandey, and P.~Tiwari, ``Sound classification using convolutional neural network and tensor deep stacking network,'' \emph{IEEE Access}, vol.~7, pp. 7717--7727, 2019.

\bibitem{chun19}
I.~Y. Chun and J.~A. Fessler, ``Convolutional analysis operator learning: Acceleration and convergence,'' \emph{IEEE Trans. Image Process.}, vol.~28, no.~11, pp. 5323--5337, 2019.

\bibitem{FISTA}
A.~Beck and M.~Teboulle, ``A fast iterative shrinkage-thresholding algorithm for linear inverse problems,'' \emph{SIAM J. Imag. Sci.}, vol.~2, no.~1, pp. 183--202, 2009.

\bibitem{BV04}
S.~Boyd and L.~Vandenberghe, \emph{Convex Optimization}.\hskip 1em plus 0.5em minus 0.4em\relax USA: Cambridge University Press, 2004.

\bibitem{CCS10}
J.-F. Cai, E.~J. Candès, and Z.~Shen, ``A singular value thresholding algorithm for matrix completion,'' \emph{SIAM J. Optim.}, vol.~20, no.~4, pp. 1956--1982, 2010.

\bibitem{K2015}
K.~J. Piczak, ``Esc: Dataset for environmental sound classification,'' in \emph{Proc. of the 23rd ACM Int. Conf. on Multimedia}, Oct. 2015, pp. 1015--1018.

\bibitem{SZ2015}
K.~Simonyan and A.~Zisserman, ``Very deep convolutional networks for large-scale image recognition,'' in \emph{3rd Int. Conf. on Learning Representations, {ICLR} 2015, San Diego, CA, USA, May 7-9, 2015, Conf. Track Proc.}, 2015.

\end{thebibliography}

\end{document}